\documentclass[prb,twocolumn,superscriptaddress,secnumarabic,balancelastpage,amsmath,amssymb,nofootinbib]{revtex4-1}
\usepackage{amssymb}
\usepackage{amsmath}
\usepackage{lgrind}        % convert program listings to a form includable in a LaTeX document
\usepackage{color}         % produces boxes or entire pages with colored backgrounds
\usepackage{graphics}      % standard graphics specifications
\usepackage[pdftex]{graphicx}      % alternative graphics specifications
\usepackage{longtable}     % helps with long table options
\usepackage{epsf}          % old package handles encapsulated post script issues
\usepackage{bm}            % special 'bold-math' package
\usepackage{asymptote}     % For typesetting of mathematical illustrations
\usepackage{thumbpdf}
\usepackage{verbatim}
\usepackage{url}
\usepackage[colorlinks=true]{hyperref}  % this package should be added after all others
\usepackage{enumitem}	% compact enumeration
\usepackage{float}
\usepackage{times}
\usepackage{color}
\usepackage{verbatim}

\begin{document}

% Include your paper's title here

\title{Exceptional Surfaces in $\mathcal{PT}$-Symmetric Photonic Systems}

% Place the author information here.  Please hand-code the contact
% information and notecalls; do *not* use \footnote commands.  Let the
% author contact information appear immediately below the author names
% as shown.  We would also prefer that you don't change the type-size
% settings shown here.

\author{Hengyun Zhou}
\affiliation{Department of Physics, Harvard University, Cambridge, Massachusetts 02138, USA}
\author{Jong Yeon Lee}
\affiliation{Department of Physics, Harvard University, Cambridge, Massachusetts 02138, USA}
\author{Shang Liu}
\affiliation{Department of Physics, Harvard University, Cambridge, Massachusetts 02138, USA}
\author{Bo Zhen}
\affiliation{Department of Physics and Astronomy, University of Pennsylvania, Philadelphia, Pennsylvania 19104, USA}

% Include the date command, but leave its argument blank.

\begin{abstract}
Exceptional points in non-Hermitian systems have recently been shown to possess nontrivial topological properties, and to give rise to many exotic physical phenomena. However, most studies thus far have focused on isolated exceptional points or one-dimensional lines of exceptional points. Here, we substantially expand the space of exceptional systems by designing two-dimensional surfaces of exceptional points, and find that symmetries are a key element to protect such exceptional surfaces. We construct them using symmetry-preserving non-Hermitian deformations of topological nodal lines, and analyze the associated symmetry, topology, and physical consequences. As a potential realization, we simulate a parity-time-symmetric 3D photonic crystal and indeed find the emergence of exceptional surfaces. Our work paves the way for future explorations of systems of exceptional points in higher dimensions.
\end{abstract}
\maketitle

\section{Introduction}
In recent years, there has been growing interest in exploring novel effects in non-Hermitian physical systems \cite{moiseyev2011non,heiss2012physics,el-ganainy2018non}. This has provided great insight into aspects of fundamental science, including parity-time ($\mathcal{PT}$) symmetry \cite{bender1997real,makris2008beam,guo2009observation,ruter2010observation} and novel topological phases \cite{alvarez2018topological,rudner2009topological,esaki2011edge,yuce2015topological,lee2016anomalous,leykam2017edge,shen2018topological,gong2018topological,lieu2018topological,lieu2018topological1,kunst2018biorthogonal,kawabata2018non,yao2018non,yao2018edge,xiong2018why,lee2018anatomy,carlstrom2018exceptional,dembowski2001experimental,poli2015selective,zeuner2015observation,weimann2016topologically,zhou2018observation,cerjan2018experimental}, while also opening the doors to a host of new applications, such as unconventional transmission and reflection \cite{lin2011unidirectional,feng2013experimental}, sensing functionalities \cite{chen2017exceptional,hodaei2017enhanced,zhang2018quantum,lau2018non}, novel lasers \cite{hodaei2014parity,feng2014single,peng2014loss,harari2018topological,bandres2018topological}, and chiral mode transfer \cite{xu2016topological,doppler2016dynamically}. Key to many of these explorations are the special properties of exceptional points (EPs)---unique spectral degeneracies where the real and imaginary parts of two or more eigenvalues coincide and the eigenvectors coalesce \cite{rotter2009non,heiss2012physics,moiseyev2011non}. Indeed, only realizable in non-Hermitian systems, EPs mark the boundaries of $\mathcal{PT}$ phase transitions and give rise to the unconventional optical responses mentioned above. Moreover, they have been shown to possess topological properties such as a $\pi$ Berry phase and vorticity \cite{dembowski2001experimental,mailybaev2005geometric,shen2018topological}, and can give rise to open Fermi arcs in the bulk dispersion of systems \cite{zhou2018observation,kozii2017non}.

However, the majority of studies thus far have focused on the properties of isolated points (0D) \cite{liertzer2012pump,xu2016topological,doppler2016dynamically,zhou2018observation} or continuous lines (1D) \cite{zhen2015spawning,cerjan2016exceptional,xu2017weyl,cerjan2018effects,cerjan2018experimental} of EPs. Correspondingly, this has limited the types of achievable band dispersions and observable phenomena in these non-Hermitian systems. This calls for approaches to go beyond these lower dimensional EP systems and explore higher dimensional configurations, such as surfaces of EPs.

In this paper, we propose and analyze several models to realize exceptional surfaces. First, we examine the general conditions for EPs to occur, and find that the emergence of EP surfaces require additional symmetry protection, indicating that EP surfaces can be understood as the generalization of Hermitian symmetry-protected nodal phases \cite{fang2016topological,chiu2016classification,armitage2018weyl,burkov2011topological,fang2015topological,zhang2016quantum,bian2016topological} to the non-Hermitian setting. Motivated by this, we consider a non-Hermitian deformation of topological nodal lines, and show that under certain symmetry conditions, this can give rise to a non-Hermitian system with EPs configured in a torus geometry. We show that the EP torus is characterized by separated components inside and outside the torus, as well as a quantized non-Hermitian Berry phase inherited from the Hermitian nodal line. In addition, we find that the EP surface encloses an open nodal volume---a three-dimensional generalization of bulk Fermi arcs---in which the real part of two bands are degenerate with each other within an entire volume. This offers remarkable control of the band structure and spectral density of states of  the system, which could have interesting applications such as enhancing nonlinearities in optical systems. Finally, we utilize $\mathcal{PT}$-symmetric gain-loss modulation in a simple three-dimensional photonic crystal structure to realize EP surfaces, which can be readily implemented in experiments.

\section{Dimensionality of EP Configurations} 
We start by considering the conditions to create EPs, in order to understand why 0D or 1D EP configurations are typically generated. Exceptional points typically occur when the real and imaginary parts of two or more eigenvalues coalesce and the eigenspace becomes defective. Mathematically, we can consider a generic non-Hermitian Dirac Hamiltonian
\begin{align}
H=\sum_{i=1}^d c_i(\vec{k})\gamma_i + c_0(\vec{k}),
\end{align}
where $\vec{k}$ denotes the momentum in dimension $d$, the Hermitian matrices $\gamma_i$ obey anti-commutation relations $\{\gamma_i,\gamma_j\}=2\delta_{i,j}$, and the functions $c_i(\vec{k})$ are complex coefficients that include the non-Hermitian nature of the system. The eigenvalues will thus take the generic form $c_0(\vec{k})\pm\sqrt{\sum_i c_i(\vec{k})^2}$, and the desired band degeneracy typically occurs when the argument of the square root equals zero: i.e. both the real and imaginary parts of the argument vanish.

Thus, we shall generically find that two constraints need to be satisfied to create an EP. The dimensionality of the EP contour generated will then be $d-2$ in spatial dimension $d$. Indeed, this result has been corroborated in various studies, where in 2D, single EPs or pairs of EPs have been observed \cite{liertzer2012pump,xu2016topological,doppler2016dynamically,zhou2018observation}, and in 3D, lines of EPs have been proposed \cite{xu2017weyl,cerjan2018effects} and observed \cite{cerjan2018experimental}. However, this also implies that additional mechanisms are required to realize EP surfaces in physical dimensions within 3.

At this stage, it is instructive to examine the case of Hermitian systems: in 3D, the robust spectral degeneracies are 0D Weyl points \cite{wan2011topological,xu2015discovery,lv2015experimental,lu2015experimental,armitage2018weyl}, and the realization of 1D nodal lines requires symmetry protection \cite{burkov2011topological,fang2016topological}. In analogy, we also expect the 2D EP surfaces to be symmetry-protected. The symmetry conditions, which can include crystalline symmetries or more general non-Hermitian symmetries \cite{bernard2001,magnea2008random,lieu2018topological1}, can reduce the number of independent equations that need to be simultaneously satisfied, leading to a $d-1$ dimensional EP configuration in $d$ dimensions. Indeed, the known examples in the literature where $d-1$ dimensional EP configurations are found, such as the conventional $\mathcal{PT}$-symmetry breaking transition that describes a 0D EP in 1D \cite{bender1997real}, and the 1D exceptional rings \cite{zhen2015spawning} and exceptional contours \cite{cerjan2016exceptional} in 2D, are protected by $\mathcal{PT}$ symmetry. Similarly, we expect $\mathcal{PT}$ symmetry to be sufficient to generate EP surfaces in 3D. We note that adding a constant gain/loss term will shift the spectrum and nominally break the $\mathcal{PT}$ symmetry, but the spectral degeneracies will remain unchanged, and thus such a shifted $\mathcal{PT}$-symmetric will also host EP surfaces.

\begin{figure}[htbp]
\centering
\includegraphics[width=\linewidth]{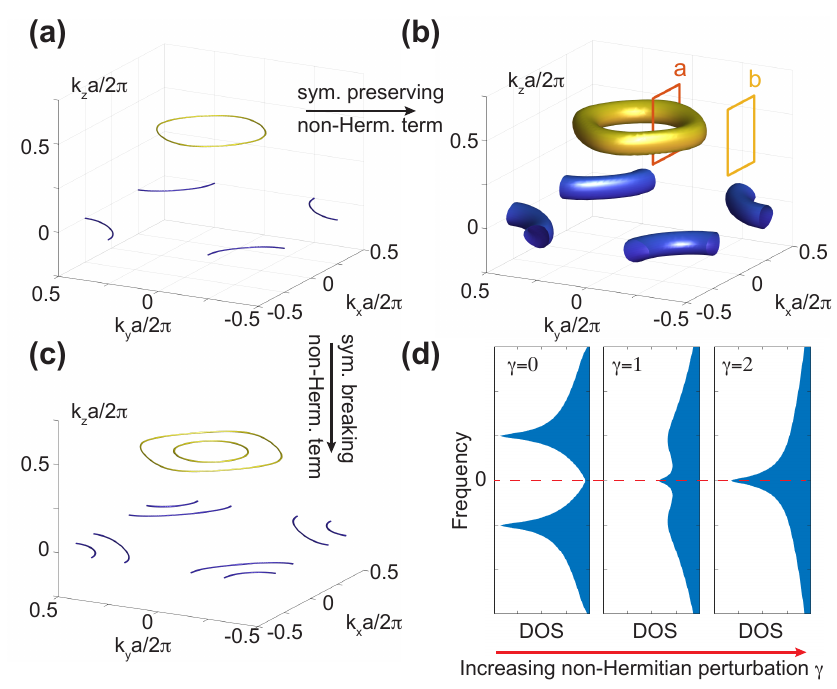}
\caption{(a,b,c) Spectral degeneracies in Hermitian and non-Hermitian systems from the model in Eq.~(\ref{eq:twoband}). The Hermitian nodal line (a) is protected by a $\mathcal{PT}$ symmetry. A symmetry-preserving non-Hermitian perturbation produces an exceptional torus (b), while a symmetry-breaking one produces exceptional rings (c). (d) Spectral density of states for increasing non-Hermitian perturbation $\gamma$, with the density of states close to the nodal line frequency (red dashed line) increasing significantly.}
\label{fig:fig2}
\end{figure}

\begin{figure}[htbp]
\centering
\includegraphics[width=\linewidth]{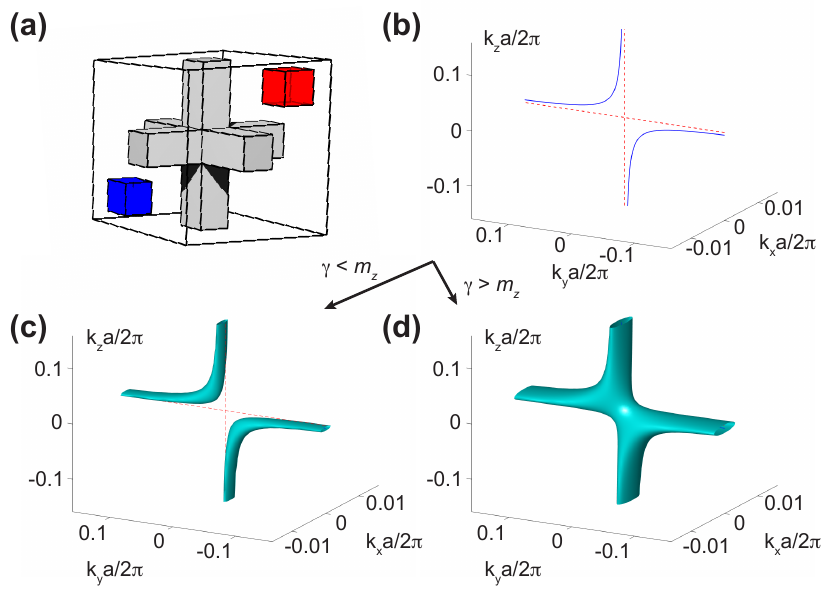}
\caption{(a) Metallic mesh structure with $\mathcal{PT}$-symmetric perturbation to realize a nodal chain in the Hermitian limit and EP surface under non-Hermitian perturbations. (b) Nodal chain crossing point (red dashed lines) can be gapped out by a mass term (blue solid lines).  Depending on the non-Hermitian perturbation strength, the gap between the EP tori can remain (c) or become connected (d).}
\label{fig:fig3}
\end{figure}

\section{EP Torus from a Topological Nodal Line}
Motivated by the fact that both Hermitian nodal lines and EP surfaces can be protected by $\mathcal{PT}$-symmetry, we now consider symmetry-preserving non-Hermitian deformations to Hermitian nodal lines. This approach thus directly makes clear the connection between EP surfaces---examples of non-Hermitian nodal phases---and their Hermitian counterparts. Compared to previous work that has studied non-Hermitian nodal lines \cite{wang2018non}, the symmetry preservation plays a key role in generating EP surfaces as opposed to lines of EPs.

First, we consider a minimal two-band model describing an EP surface, with the Hamiltonian given by
\begin{align}
H(\vec{k})&=(m-6+2\cos(k_x)+2\cos(k_y)+2\cos(k_z))\sigma_z \nonumber\\&+ 2\lambda\sin(k_z)\sigma_x+i\gamma\sigma_y+E_0-i\gamma_0,
\label{eq:twoband}
\end{align}
where the non-Hermiticity enters through the $\sigma_y$ term, and the remaining terms describe a conventional Hermitian nodal ring, shifted by energy $E_0$ and loss $\gamma_0$. In the Hermitian limit, the nodal line exists for $m>0$, but shrinks as $m$ is decreased and vanishes for $m<0$.

For illustration purposes, we choose parameters $m=6$, $\lambda=1$. In the Hermitian limit $\gamma=0$, the spectral degeneracies of the system are shown in Fig.~\ref{fig:fig2}(a), clearly showing two nodal rings. Once a non-Hermitian perturbation $\gamma=0.8$ is included, each nodal ring splits into a torus of EPs, as shown in Fig.~\ref{fig:fig2}(b). The non-Hermitian perturbation is specifically chosen to respect the combined $\mathcal{PT}$ symmetry of the nodal line, where $P=\sigma_z$ and $T=\sigma_z\mathcal{K}$, $\mathcal{K}$ being the complex-conjugation operation. This guarantees that in the non-Hermitian setting, the only symmetry-admissible perturbation to the nodal line Hamiltonian is proportional to $i\sigma_y$, thus resulting in an EP torus (we do not impose the symmetry on the shift proportional to identity, as it does not affect the spectral degeneracies). On the other hand, if we choose a perturbation that breaks the $\mathcal{PT}$-symmetry, e.g. $i\gamma\sigma_x$ instead of $i\gamma\sigma_y$, then the conditions for EPs to occur become $k_z=0/\pi$, $m-4+2\cos(k_x)+2\cos(k_y)=\pm \gamma$, which describe four EP rings instead of EP surfaces \cite{wang2018non}, as shown in Fig.~\ref{fig:fig2}(c).

The physics discussed here can be easily realized in microwave experiments, for example in the system employed in Ref.~\onlinecite{yan2018experimental}, which consists of a metallic-mesh 3D photonic crystal. Nodal chains have been discovered in this simple structure, where the nodal lines are protected by the combined $\mathcal{PT}$-symmetry action and the chain crossing point is additionally protected by a mirror symmetry. 

By adding non-Hermitian perturbations that preserve the $\mathcal{PT}$-symmetry, as shown in Fig.~\ref{fig:fig3}(a), each nodal line can be split into an EP surface. The deformation to the chain crossing point however can be more complicated, and in general may lead to novel EP geometries. At the level of an effective two-band Hamiltonian, consider the local Hamiltonian in $\vec{k}$-space in the vicinity of the nodal chain crossing point: 
\begin{align}
H(\vec{k})=k_x\sigma_x+(k_yk_z+m_z)\sigma_z+i\gamma\sigma_y,
\end{align}where the mass term $m_z$ and non-Hermitian perturbation $\gamma$ can appear when the mirror symmetry protecting the chain crossing point is broken. Although the presence of the mass term will gap out the Hermitian nodal chain crossing point (Fig.~\ref{fig:fig3}(b)), a non-Hermitian term that is larger in magnitude can restore the connectedness of the feature (Fig.~\ref{fig:fig3}(d)), even if the mirror symmetry of the system is now broken.

\begin{figure}[htbp]
\centering
\includegraphics[width=\linewidth]{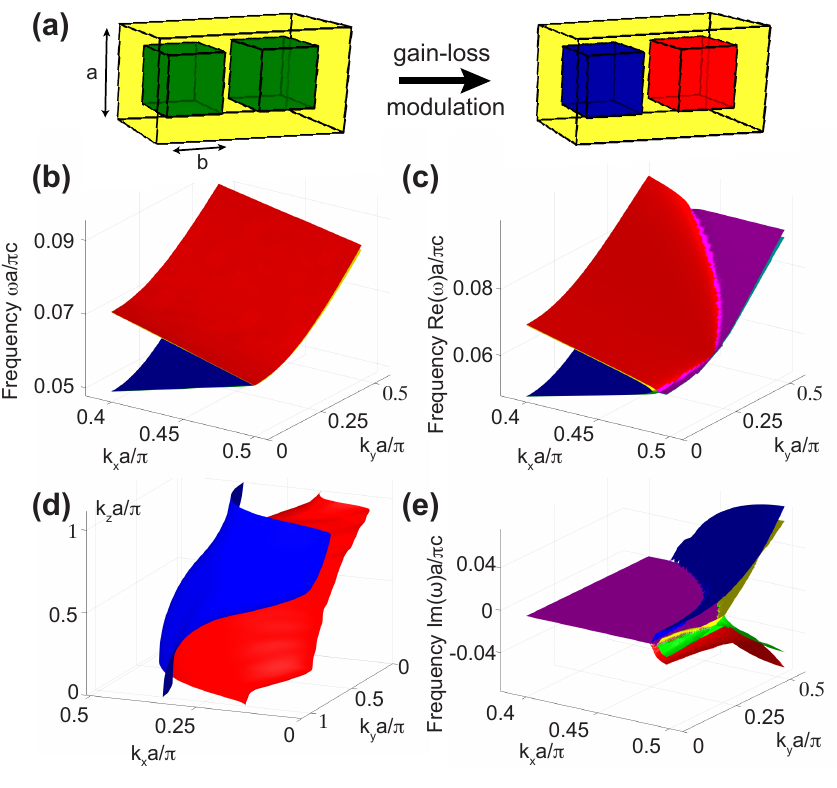}
\caption{(a) A photonic crystal consisting of a cubic array (lattice constant $a=1$) of dielectric cubes (side length $b=0.6a$), with dielectric constant $\epsilon=12$. Under gain-loss modulation, where the blue (red) cube experiences a loss (gain) $\gamma=10$, an exceptional surface appears. (b) 2D cross-section of the band structure near the band-folding line, at a generic $k_z=0.13\pi/a$, in the Hermitian limit $\gamma=0$. (c,e) Similar 2D cross section of the real (c) and imaginary (e) parts of the band structure in the non-Hermitian modulated case $\gamma=10$. (d) Extracted EP surfaces for the model in (c,e).}
\label{fig:fig1}
\end{figure}

\section{Topological and Physical Properties of EP Surfaces}
We now turn to analyze the topological properties associated with the EP torus obtained above. For a Hamiltonian with $\mathcal{PT}$-symmetry, we can choose a basis such that the $\mathcal{PT}$-symmetry corresponds to complex-conjugation, in which case the Hamiltonian is guaranteed to be real, and complex eigenvalues must come in complex-conjugate pairs. Therefore, when changing a Hamiltonian in which all eigenvalues are real into one in which there are complex-conjugate pairs, the continuity of eigenvalues requires a degeneracy of eigenvalues on the real line to appear in between, corresponding to an EP. Thus, the inside (complex eigenvalues) and outside (real eigenvalues) of the torus belong to two disconnected branches, and there must exist a continuous surface of EPs in between them. This is further characterized by the discriminant of the characteristic polynomial of the Hamiltonian (see appendix).

In addition, we find that the generalization of the Hermitian Berry phase protects the EP surface to form a continuous torus. Since the Hamiltonian still respects $\mathcal{PT}$-symmetry in the non-Hermitian case, we shall find that the Berry phase $\varphi_{RR}=-i\oint_c\langle \psi_R(\vec{k})|\nabla_{\vec{k}}| \psi_R(\vec{k})\rangle$, defined via right eigenvectors on both sides, is still quantized to $0$ or $\pi$ (mod $2\pi$), as long as the integration path chosen is in the $\mathcal{PT}$-unbroken phase \cite{bender1997real}, where all eigenvalues are real (see appendix for details). Note that by continuity, the region outside of the EP torus, which was originally gapped in the Hermitian limit, will belong to the $\mathcal{PT}$-unbroken phase and have real eigenvalues. However, the Berry phase in the interior of the torus is no longer quantized.

For the preceding model in Eq.~(\ref{eq:twoband}), we calculate the Berry phase along different paths outside the  torus using the normalized eigenvectors. For a path that does not link with the EP torus (Fig.~\ref{fig:fig2}(b), path [b]), the Berry phase is trivial, while along a path that links with the EP torus (Fig.~\ref{fig:fig2}(b), path [a]), we find that the Berry phase is equal to $\pi$. These results are consistent with the $\pi$ Berry phase for the Hermitian nodal line, indicating that this Hermitian topological invariant is inherited by the non-Hermitian system, and correspondingly the topological protection against the EP torus breaking apart remains.

The generation of EP surfaces also has important consequences for the bulk dispersion of the system. Similar to the bulk Fermi arcs that terminate at discrete EPs \cite{kozii2017non,shen2018topological,zhou2018observation}, the internal volume of the EP surface will have a pair of bands that are completely degenerate in the real part of their eigenvalues, which we call an open nodal volume. The non-Hermitian term then provides remarkable control of the spectral density of states of the system: as we continuously increase the non-Hermitian perturbation from 0 to a large value, the density of states near the nodal line frequency will be continuously tuned from a relatively small value, due to the 1D nature of the nodal line, to a large value, due to the 3D volumetric factor of the open nodal volume. We calculate the spectral density of states for the model Eq.~(\ref{eq:twoband}), taking into account the imaginary part of the energy as a Lorentzian width. In view of experiments, we have added a constant loss term $\gamma_0=\gamma+0.1$ to make the system completely passive. The results are illustrated in Fig.~\ref{fig:fig2}(d), where the spectral density of states near the nodal line clearly increases as the strength of the non-Hermitian perturbation is increased.

In general, ${\cal P T}$ symmetry allows a $k$-dependent frequency shift $E_0(k)$ in Eq.~(\ref{eq:twoband}) instead of the constant $E_0$. Thus, in physical realizations of nodal lines and the corresponding EP surfaces, the nodal line may not be completely flat in frequency, which will cause some reduction of the accessible range of density of states. However, the structure shown in Fig.~\ref{fig:fig3}(a) has been shown to have a remarkably uniform nodal line frequency in the Hermitian limit, with variations less than $1\%$ across the whole Brillouin zone. This makes it a promising system to observe the EP torus discussed here, as well as to investigate the evolution of density of states with the strength of the non-Hermitian term. In addition, the system can be engineered to possess a non-Hermitian particle-hole symmetry \cite{qi2018defect}, which can protect the nodal volume to be completely flat.

\section{EP Surfaces in a PT-Symmetric Photonic Crystal}
We now consider a concrete realization of EP surfaces in a $\mathcal{PT}$-symmetric photonic crystal, obtained from gain-loss modulation in a regular 3D dielectric photonic crystal. As shown in Fig.~\ref{fig:fig1}(a), the Hermitian photonic crystal consists of dielectric cubes arranged in a cubic lattice, with parameters labeled in the figure and captions. The photonic band dispersion in a 2D cut at a generic $k_z$ value of this structure is shown in Fig.~\ref{fig:fig1}(b), where along the $k_y=0$ axis, each colored band actually consists of two degenerate bands with different polarizations. By doubling the unit cell size in one spatial direction, and applying loss on one site and gain on the next, we form a $\mathcal{PT}$-symmetric photonic crystal with a super-cell size twice that of the original system.

In Fig.~\ref{fig:fig1}(c) and Fig.~\ref{fig:fig1}(e), we show the same 2D cut of the real and imaginary parts of the resulting band structure when the $\mathcal{PT}$-symmetric non-Hermitian perturbation is applied. We find that the band structure is separated into two regions, one in which the real part of multiple bands coalesce, and one in which the imaginary parts coalesce. The surface separating these two regions is composed of two bands completely degenerate in their real and imaginary part, and thus corresponds to a surface of EPs. By extracting the locations at which this occurs, we find in Fig.~\ref{fig:fig1}(d) that there are indeed two EP surfaces in the band structure. The two EP surfaces originate from different polarizations.

While we have chosen a relatively large non-Hermitian term to make the effects more apparent, we note that due to the initial band-folding degeneracy after doubling the unit cell, a threshold-less $\mathcal{PT}$ transition is realized \cite{cerjan2016exceptional}, and any finite amount of non-Hermitian modulation is in fact sufficient to produce an EP surface. The resulting band structure can be quite significantly modified by the non-Hermitian terms, showing relatively flat real part dispersions in the direction perpendicular to the band-folding line (Fig.~\ref{fig:fig1}(c)) and sharp changes in the gain/loss response as the $\vec{k}$-vector is tuned (Fig.~\ref{fig:fig1}(e)). In addition, a constant shift in complex energy will not affect the EP surface, so the same design can be implemented in purely passive systems as well.

\section{Conclusion and Discussion} 
To conclude, in this work, we have proposed various methods to realize EP surfaces, analyzed their symmetry and topology, and discussed straightforward avenues to their experimental implementation in photonic and microwave systems. We have also shown that the bulk dispersion of these systems is drastically modified by the inclusion of symmetry-preserving non-Hermitian terms, giving rise to a highly tunable spectral density of states.

While our discussion has focused on systems with $\mathcal{PT}$-symmetry, the analysis can be readily generalized to other types of symmetries, including both the conventional symmetries specifying the AZ classes \cite{altland1997nonstandard,chiu2016classification} and crystalline symmetries, as well as more general types of symmetries according to the Bernard-LeClair classification \cite{bernard2001,magnea2008random,lieu2018topological1,zhou2018periodic}. Similarly, it may be interesting to consider other types of invariants that may be defined in this system (such as those corresponding to the second homotopy group), particularly those that are unique to non-Hermitian systems. In addition, although we have been focusing on the bulk properties of EP surfaces, it may be interesting to consider their realization in systems of finite extent, and understand the associated bulk-boundary correspondence. Indeed, it has been found in lower-dimensional models that the conventional bulk-boundary correspondence must be significantly modified, and unusual phenomena such as the non-Hermitian skin effect can emerge \cite{kunst2018biorthogonal,yao2018non,yao2018edge,xiong2018why,lee2018anatomy}.

\begin{acknowledgements}
We acknowledge helpful discussions with M.~Solja\v{c}i\'{c}, M.~Lukin, D.~Borgnia, L.~Chen, Y.~Fu and N.~Rivera.

Note added: towards the completion of this project, we became aware of related contributions by J.~Budich et al. \cite{budich2018symmetry}, and R.~Okugawa et al. \cite{okugawa2018topological}, where exceptional surfaces in non-Hermitian systems are also explored.
\end{acknowledgements}

\appendix
\section{Details of Topological Invariants}
In this appendix, we present more details for the calculation of topological invariants for the EP torus with $\mathcal{PT}$-symmetry.

We can choose a basis such that the combined $\mathcal{PT}$ operation, being an anti-linear, anti-unitary symmetry that squares to 1, maps $H(\vec{k})\rightarrow H^*(\vec{k})$. This implies that the Hamiltonian can be chosen to be real at each point, and that the complex eigenvalues in the spectra must consist of complex conjugate pairs. As discussed in the main text, this implies, by continuity, that the emergence of complex conjugate pairs in the spectra from real eigenvalues necessarily involves passing through a degeneracy in eigenvalues.

More generally, for a Hamiltonian $H(\vec{k})$, we may write down the characteristic polynomial. The condition for a spectral degeneracy to occur (i.e. the Hamiltonian has degenerate eigenvalues) is that the discriminant of the characteristic polynomial vanishes. At a $\vec{k}$ point where the spectrum is completely gapped, we can calculate the sign of the discriminant of the characteristic polynomial (this is a real number because the matrix is constrained to be real); positive and negative values must then necessarily belong to different disconnected components, and an EP surface is guaranteed to occur between them. This is a topological invariant that is of the type of a zeroth homotopy.

As an example, we consider the case of two bands with a real Hamiltonian, such that
\begin{align}
H=c_1\sigma_x+c_2(i\sigma_y)+c_3\sigma_z+c_4,
\end{align}
where $c_i$ are real numbers. The characteristic polynomial is
\begin{align}
|\lambda I-H|=(\lambda-c_4)^2-c_3^2-c_1^2+c_2^2,
\end{align}
and the discriminant of the characteristic polynomial is
\begin{align}
\textrm{discrim}(|\lambda  I-H|)=4(c_3^2+c_1^2-c_2^2).
\end{align}
In the concrete model described in Eq.~(\ref{eq:twoband}), for a point inside the EP torus, such as a point on the original Hermitian nodal line, $c_1=c_3=0$, $c_2=\gamma$, so $\textrm{sgn}(\textrm{discrim}(|\lambda  I-H|))=-4\gamma^2<0$. Meanwhile, outside the EP torus, such as at the origin when $\gamma$ is small, $c_1=0$, $c_2=\gamma$, $c_3=m$, $\textrm{sgn}(\textrm{discrim}(|\lambda  I-H|))=4(m^2-\gamma^2)>0$. Thus, the two points lie in different topological components, and there must be an EP surface separating them.

To generalize the Berry phase of the Hermitian nodal line to the non-Hermitian setting, we utilize the non-Hermitian Berry phase defined with right eigenvectors $|\psi_R(\vec{k})\rangle$:
\begin{align}
\phi_{RR}=-i\oint_c \langle\psi_R(\vec{k})|\nabla_{\vec{k}}|\psi_R(\vec{k})\rangle d\vec{k}.
\end{align}
$\phi_{RR}$ is defined up to $2\pi$, since we can multiply a continuously varying $k$-dependent phase factor in front of the Bloch states.

In the $\mathcal{PT}$-unbroken phase, where all the eigenvalues are real and non-degenerate, the presence of ${\cal PT}$ symmetry implies that $\mathcal{PT}|\psi_R(\vec{k})\rangle=\textrm{e}^{i\theta_{\vec{k}}}|\psi_R(\vec{k})\rangle$. Since the eigenvalues are real, the condition that the path is gapped in the complex plane implies that there is no band-switching on the path, and the phase $\theta_{\vec{k}}$ can be chosen to be continuous. Thus, we have ${\cal PT} \phi_{RR} \equiv \phi_{RR} \,\, \textrm{mod }2\pi$. On the other hand, we have
\begin{align}
{\cal PT} \phi_{RR}&=-i\oint_c \langle\mathcal{PT}\psi_R(\vec{k})|\nabla_{\vec{k}}|\mathcal{PT}\psi_R(\vec{k})\rangle d\vec{k}\nonumber\\
&=-i \oint_c  ( \nabla_{\vec{k}}\langle\psi_R(\vec{k})| )|\psi_R(\vec{k})\rangle d\vec{k}\nonumber\\
&= i\oint_c \langle\psi_R(\vec{k})| ( \nabla_{\vec{k}}|\psi_R(\vec{k}) \rangle )d\vec{k}  \nonumber\\
&=-\phi_{RR},
\end{align}
where we used that ${\cal PT}$ is an anti-unitary operator on the second line, and integration by part on the third line. Equating the two results, we have $\phi_{RR} \equiv -\phi_{RR} \,\, \textrm{mod }2\pi$. Thus, the non-Hermitian Berry phase defined with right eigenvectors is quantized to be 0 or $\pi$ modulo $2\pi$. 

We note that this relation does not hold in the $\mathcal{PT}$-broken phase, since the eigenvalues in this case come in complex conjugate pairs, and the $\mathcal{PT}$ operation can map an eigenstate to its complex conjugate pair. However, the paths that we choose for the integration, which are outside the EP torus, are guaranteed to be in the $\mathcal{PT}$-unbroken phase, since they are non-degenerate in the Hermitian limit and have not yet passed through an EP as the non-Hermitian term is increased.

We can explicitly calculate the non-Hermitian Berry phase along a loop linking with the EP torus. Considering a path located on a 2D cut $k_x=0$ of the torus, and writing down a local model in this 2D plane near the original nodal line, the Hamiltonian can be written as (after a basis change)
\begin{align}
H_{2D}(\vec{k})&=k_y\sigma_x+k_z\sigma_y+i\gamma\sigma_z\nonumber\\&=\sqrt{k^2-\gamma^2}\begin{pmatrix}
i\sinh\theta & \cosh\theta\textrm{e}^{-i\varphi}\\
\cosh\theta\textrm{e}^{i\varphi} & -i\sinh\theta\\
\end{pmatrix},
\end{align}
where we have defined $k=\sqrt{k_y^2+k_z^2}$, $\varphi=\textrm{arg}(k_y+ik_z)$, $\sinh\theta=\gamma/\sqrt{k^2-\gamma^2}$, $\cosh\theta=k/\sqrt{k^2-\gamma^2}$, where we have chosen a loop with $k>\gamma$ such that the loop is completely outside the EP torus (and hence in the $\mathcal{PT}$-unbroken phase) and links with the torus. The normalized eigenvectors can then be written as
\begin{align}
|\psi_\pm\rangle=\frac{1}{\sqrt{2}}\begin{pmatrix}
\textrm{e}^{-i\varphi}\\ \frac{\pm 1-i\sinh\theta}{\cosh\theta}
\end{pmatrix}.
\end{align}
Integrating using the preceding definition of the non-Hermitian Berry phase then gives a value of $\pi$. On the other hand, for a path that is in the $\mathcal{PT}$-broken phase, we find that the Berry phase is no longer quantized.

\bibliography{main}

\begin{thebibliography}{71}%
\makeatletter
\providecommand \@ifxundefined [1]{%
 \@ifx{#1\undefined}
}%
\providecommand \@ifnum [1]{%
 \ifnum #1\expandafter \@firstoftwo
 \else \expandafter \@secondoftwo
 \fi
}%
\providecommand \@ifx [1]{%
 \ifx #1\expandafter \@firstoftwo
 \else \expandafter \@secondoftwo
 \fi
}%
\providecommand \natexlab [1]{#1}%
\providecommand \enquote  [1]{``#1''}%
\providecommand \bibnamefont  [1]{#1}%
\providecommand \bibfnamefont [1]{#1}%
\providecommand \citenamefont [1]{#1}%
\providecommand \href@noop [0]{\@secondoftwo}%
\providecommand \href [0]{\begingroup \@sanitize@url \@href}%
\providecommand \@href[1]{\@@startlink{#1}\@@href}%
\providecommand \@@href[1]{\endgroup#1\@@endlink}%
\providecommand \@sanitize@url [0]{\catcode `\\12\catcode `\$12\catcode
  `\&12\catcode `\#12\catcode `\^12\catcode `\_12\catcode `\%12\relax}%
\providecommand \@@startlink[1]{}%
\providecommand \@@endlink[0]{}%
\providecommand \url  [0]{\begingroup\@sanitize@url \@url }%
\providecommand \@url [1]{\endgroup\@href {#1}{\urlprefix }}%
\providecommand \urlprefix  [0]{URL }%
\providecommand \Eprint [0]{\href }%
\providecommand \doibase [0]{http://dx.doi.org/}%
\providecommand \selectlanguage [0]{\@gobble}%
\providecommand \bibinfo  [0]{\@secondoftwo}%
\providecommand \bibfield  [0]{\@secondoftwo}%
\providecommand \translation [1]{[#1]}%
\providecommand \BibitemOpen [0]{}%
\providecommand \bibitemStop [0]{}%
\providecommand \bibitemNoStop [0]{.\EOS\space}%
\providecommand \EOS [0]{\spacefactor3000\relax}%
\providecommand \BibitemShut  [1]{\csname bibitem#1\endcsname}%
\let\auto@bib@innerbib\@empty
%</preamble>
\bibitem [{\citenamefont {Moiseyev}(2011)}]{moiseyev2011non}%
  \BibitemOpen
  \bibfield  {author} {\bibinfo {author} {\bibfnamefont {N.}~\bibnamefont
  {Moiseyev}},\ }\href
  {https://books.google.com/books/about/Non{\_}Hermitian{\_}Quantum{\_}Mechanics.html?id=QwAXVvk{\_}57QC{\&}pgis=1}
  {\emph {\bibinfo {title} {{Non-Hermitian Quantum Mechanics}}}}\ (\bibinfo
  {publisher} {Cambridge University Press},\ \bibinfo {year}
  {2011})\BibitemShut {NoStop}%
\bibitem [{\citenamefont {Heiss}(2012)}]{heiss2012physics}%
  \BibitemOpen
  \bibfield  {author} {\bibinfo {author} {\bibfnamefont {W.~D.}\ \bibnamefont
  {Heiss}},\ }\href {\doibase 10.1088/1751-8113/45/44/444016} {\bibfield
  {journal} {\bibinfo  {journal} {Journal of Physics A: Mathematical and
  General}\ }\textbf {\bibinfo {volume} {45}},\ \bibinfo {pages} {444016}
  (\bibinfo {year} {2012})}\BibitemShut {NoStop}%
\bibitem [{\citenamefont {El-Ganainy}\ \emph {et~al.}(2018)\citenamefont
  {El-Ganainy}, \citenamefont {Makris}, \citenamefont {Khajavikhan},
  \citenamefont {Musslimani}, \citenamefont {Rotter},\ and\ \citenamefont
  {Christodoulides}}]{el-ganainy2018non}%
  \BibitemOpen
  \bibfield  {author} {\bibinfo {author} {\bibfnamefont {R.}~\bibnamefont
  {El-Ganainy}}, \bibinfo {author} {\bibfnamefont {K.~G.}\ \bibnamefont
  {Makris}}, \bibinfo {author} {\bibfnamefont {M.}~\bibnamefont {Khajavikhan}},
  \bibinfo {author} {\bibfnamefont {Z.~H.}\ \bibnamefont {Musslimani}},
  \bibinfo {author} {\bibfnamefont {S.}~\bibnamefont {Rotter}}, \ and\ \bibinfo
  {author} {\bibfnamefont {D.~N.}\ \bibnamefont {Christodoulides}},\ }\href
  {\doibase 10.1038/nphys4323} {\bibfield  {journal} {\bibinfo  {journal}
  {Nature Physics}\ }\textbf {\bibinfo {volume} {14}},\ \bibinfo {pages} {11}
  (\bibinfo {year} {2018})}\BibitemShut {NoStop}%
\bibitem [{\citenamefont {Bender}\ and\ \citenamefont
  {Boettcher}(1997)}]{bender1997real}%
  \BibitemOpen
  \bibfield  {author} {\bibinfo {author} {\bibfnamefont {C.~M.}\ \bibnamefont
  {Bender}}\ and\ \bibinfo {author} {\bibfnamefont {S.}~\bibnamefont
  {Boettcher}},\ }\href {\doibase 10.1103/PhysRevLett.80.5243} {\bibfield
  {journal} {\bibinfo  {journal} {Physical Review Letters}\ }\textbf {\bibinfo
  {volume} {80}},\ \bibinfo {pages} {5243} (\bibinfo {year} {1997})},\ \Eprint
  {http://arxiv.org/abs/9712001} {arXiv:9712001 [physics]} \BibitemShut
  {NoStop}%
\bibitem [{\citenamefont {Makris}\ \emph {et~al.}(2008)\citenamefont {Makris},
  \citenamefont {El-Ganainy}, \citenamefont {Christodoulides},\ and\
  \citenamefont {Musslimani}}]{makris2008beam}%
  \BibitemOpen
  \bibfield  {author} {\bibinfo {author} {\bibfnamefont {K.~G.}\ \bibnamefont
  {Makris}}, \bibinfo {author} {\bibfnamefont {R.}~\bibnamefont {El-Ganainy}},
  \bibinfo {author} {\bibfnamefont {D.~N.}\ \bibnamefont {Christodoulides}}, \
  and\ \bibinfo {author} {\bibfnamefont {Z.~H.}\ \bibnamefont {Musslimani}},\
  }\href {\doibase 10.1103/PhysRevLett.100.103904} {\bibfield  {journal}
  {\bibinfo  {journal} {Physical Review Letters}\ }\textbf {\bibinfo {volume}
  {100}},\ \bibinfo {pages} {103904} (\bibinfo {year} {2008})}\BibitemShut
  {NoStop}%
\bibitem [{\citenamefont {Guo}\ \emph {et~al.}(2009)\citenamefont {Guo},
  \citenamefont {Salamo}, \citenamefont {Duchesne}, \citenamefont {Morandotti},
  \citenamefont {Volatier-Ravat}, \citenamefont {Aimez}, \citenamefont
  {Siviloglou},\ and\ \citenamefont {Christodoulides}}]{guo2009observation}%
  \BibitemOpen
  \bibfield  {author} {\bibinfo {author} {\bibfnamefont {A.}~\bibnamefont
  {Guo}}, \bibinfo {author} {\bibfnamefont {G.~J.}\ \bibnamefont {Salamo}},
  \bibinfo {author} {\bibfnamefont {D.}~\bibnamefont {Duchesne}}, \bibinfo
  {author} {\bibfnamefont {R.}~\bibnamefont {Morandotti}}, \bibinfo {author}
  {\bibfnamefont {M.}~\bibnamefont {Volatier-Ravat}}, \bibinfo {author}
  {\bibfnamefont {V.}~\bibnamefont {Aimez}}, \bibinfo {author} {\bibfnamefont
  {G.~A.}\ \bibnamefont {Siviloglou}}, \ and\ \bibinfo {author} {\bibfnamefont
  {D.~N.}\ \bibnamefont {Christodoulides}},\ }\href {\doibase
  10.1103/PhysRevLett.103.093902} {\bibfield  {journal} {\bibinfo  {journal}
  {Physical Review Letters}\ }\textbf {\bibinfo {volume} {103}},\ \bibinfo
  {pages} {093902} (\bibinfo {year} {2009})}\BibitemShut {NoStop}%
\bibitem [{\citenamefont {R{\"{u}}ter}\ \emph {et~al.}(2010)\citenamefont
  {R{\"{u}}ter}, \citenamefont {Makris}, \citenamefont {El-Ganainy},
  \citenamefont {Christodoulides}, \citenamefont {Segev},\ and\ \citenamefont
  {Kip}}]{ruter2010observation}%
  \BibitemOpen
  \bibfield  {author} {\bibinfo {author} {\bibfnamefont {C.~E.}\ \bibnamefont
  {R{\"{u}}ter}}, \bibinfo {author} {\bibfnamefont {K.~G.}\ \bibnamefont
  {Makris}}, \bibinfo {author} {\bibfnamefont {R.}~\bibnamefont {El-Ganainy}},
  \bibinfo {author} {\bibfnamefont {D.~N.}\ \bibnamefont {Christodoulides}},
  \bibinfo {author} {\bibfnamefont {M.}~\bibnamefont {Segev}}, \ and\ \bibinfo
  {author} {\bibfnamefont {D.}~\bibnamefont {Kip}},\ }\href {\doibase
  10.1038/nphys1515} {\bibfield  {journal} {\bibinfo  {journal} {Nature
  Physics}\ }\textbf {\bibinfo {volume} {6}},\ \bibinfo {pages} {192} (\bibinfo
  {year} {2010})}\BibitemShut {NoStop}%
\bibitem [{\citenamefont {Alvarez}\ \emph {et~al.}(2018)\citenamefont
  {Alvarez}, \citenamefont {Vargas}, \citenamefont {Berdakin},\ and\
  \citenamefont {Torres}}]{alvarez2018topological}%
  \BibitemOpen
  \bibfield  {author} {\bibinfo {author} {\bibfnamefont {V.~M.~M.}\
  \bibnamefont {Alvarez}}, \bibinfo {author} {\bibfnamefont {J.~E.~B.}\
  \bibnamefont {Vargas}}, \bibinfo {author} {\bibfnamefont {M.}~\bibnamefont
  {Berdakin}}, \ and\ \bibinfo {author} {\bibfnamefont {L.~E. F.~F.}\
  \bibnamefont {Torres}},\ }\href {http://arxiv.org/abs/1805.08200} {\bibfield
  {journal} {\bibinfo  {journal} {arXiv}\ } (\bibinfo {year} {2018})},\ \Eprint
  {http://arxiv.org/abs/1805.08200} {arXiv:1805.08200} \BibitemShut {NoStop}%
\bibitem [{\citenamefont {Rudner}\ and\ \citenamefont
  {Levitov}(2009)}]{rudner2009topological}%
  \BibitemOpen
  \bibfield  {author} {\bibinfo {author} {\bibfnamefont {M.~S.}\ \bibnamefont
  {Rudner}}\ and\ \bibinfo {author} {\bibfnamefont {L.~S.}\ \bibnamefont
  {Levitov}},\ }\href {\doibase 10.1103/PhysRevLett.102.065703} {\bibfield
  {journal} {\bibinfo  {journal} {Physical Review Letters}\ }\textbf {\bibinfo
  {volume} {102}},\ \bibinfo {pages} {065703} (\bibinfo {year}
  {2009})}\BibitemShut {NoStop}%
\bibitem [{\citenamefont {Esaki}\ \emph {et~al.}(2011)\citenamefont {Esaki},
  \citenamefont {Sato}, \citenamefont {Hasebe},\ and\ \citenamefont
  {Kohmoto}}]{esaki2011edge}%
  \BibitemOpen
  \bibfield  {author} {\bibinfo {author} {\bibfnamefont {K.}~\bibnamefont
  {Esaki}}, \bibinfo {author} {\bibfnamefont {M.}~\bibnamefont {Sato}},
  \bibinfo {author} {\bibfnamefont {K.}~\bibnamefont {Hasebe}}, \ and\ \bibinfo
  {author} {\bibfnamefont {M.}~\bibnamefont {Kohmoto}},\ }\href {\doibase
  10.1103/PhysRevB.84.205128} {\bibfield  {journal} {\bibinfo  {journal}
  {Physical Review B}\ }\textbf {\bibinfo {volume} {84}},\ \bibinfo {pages}
  {205128} (\bibinfo {year} {2011})}\BibitemShut {NoStop}%
\bibitem [{\citenamefont {Yuce}(2015)}]{yuce2015topological}%
  \BibitemOpen
  \bibfield  {author} {\bibinfo {author} {\bibfnamefont {C.}~\bibnamefont
  {Yuce}},\ }\href {\doibase 10.1016/J.PHYSLETA.2015.02.011} {\bibfield
  {journal} {\bibinfo  {journal} {Physics Letters A}\ }\textbf {\bibinfo
  {volume} {379}},\ \bibinfo {pages} {1213} (\bibinfo {year}
  {2015})}\BibitemShut {NoStop}%
\bibitem [{\citenamefont {Lee}(2016)}]{lee2016anomalous}%
  \BibitemOpen
  \bibfield  {author} {\bibinfo {author} {\bibfnamefont {T.~E.}\ \bibnamefont
  {Lee}},\ }\href {\doibase 10.1103/PhysRevLett.116.133903} {\bibfield
  {journal} {\bibinfo  {journal} {Physical Review Letters}\ }\textbf {\bibinfo
  {volume} {116}},\ \bibinfo {pages} {133903} (\bibinfo {year}
  {2016})}\BibitemShut {NoStop}%
\bibitem [{\citenamefont {Leykam}\ \emph {et~al.}(2017)\citenamefont {Leykam},
  \citenamefont {Bliokh}, \citenamefont {Huang}, \citenamefont {Chong},\ and\
  \citenamefont {Nori}}]{leykam2017edge}%
  \BibitemOpen
  \bibfield  {author} {\bibinfo {author} {\bibfnamefont {D.}~\bibnamefont
  {Leykam}}, \bibinfo {author} {\bibfnamefont {K.~Y.}\ \bibnamefont {Bliokh}},
  \bibinfo {author} {\bibfnamefont {C.}~\bibnamefont {Huang}}, \bibinfo
  {author} {\bibfnamefont {Y.~D.}\ \bibnamefont {Chong}}, \ and\ \bibinfo
  {author} {\bibfnamefont {F.}~\bibnamefont {Nori}},\ }\href {\doibase
  10.1103/PhysRevLett.118.040401} {\bibfield  {journal} {\bibinfo  {journal}
  {Physical Review Letters}\ }\textbf {\bibinfo {volume} {118}},\ \bibinfo
  {pages} {040401} (\bibinfo {year} {2017})},\ \Eprint
  {http://arxiv.org/abs/1610.04029} {arXiv:1610.04029} \BibitemShut {NoStop}%
\bibitem [{\citenamefont {Shen}\ \emph {et~al.}(2018)\citenamefont {Shen},
  \citenamefont {Zhen},\ and\ \citenamefont {Fu}}]{shen2018topological}%
  \BibitemOpen
  \bibfield  {author} {\bibinfo {author} {\bibfnamefont {H.}~\bibnamefont
  {Shen}}, \bibinfo {author} {\bibfnamefont {B.}~\bibnamefont {Zhen}}, \ and\
  \bibinfo {author} {\bibfnamefont {L.}~\bibnamefont {Fu}},\ }\href {\doibase
  10.1103/PhysRevLett.120.146402} {\bibfield  {journal} {\bibinfo  {journal}
  {Physical Review Letters}\ }\textbf {\bibinfo {volume} {120}} (\bibinfo
  {year} {2018}),\ 10.1103/PhysRevLett.120.146402},\ \Eprint
  {http://arxiv.org/abs/1706.07435} {arXiv:1706.07435} \BibitemShut {NoStop}%
\bibitem [{\citenamefont {Gong}\ \emph {et~al.}(2018)\citenamefont {Gong},
  \citenamefont {Ashida}, \citenamefont {Kawabata}, \citenamefont {Takasan},
  \citenamefont {Higashikawa},\ and\ \citenamefont
  {Ueda}}]{gong2018topological}%
  \BibitemOpen
  \bibfield  {author} {\bibinfo {author} {\bibfnamefont {Z.}~\bibnamefont
  {Gong}}, \bibinfo {author} {\bibfnamefont {Y.}~\bibnamefont {Ashida}},
  \bibinfo {author} {\bibfnamefont {K.}~\bibnamefont {Kawabata}}, \bibinfo
  {author} {\bibfnamefont {K.}~\bibnamefont {Takasan}}, \bibinfo {author}
  {\bibfnamefont {S.}~\bibnamefont {Higashikawa}}, \ and\ \bibinfo {author}
  {\bibfnamefont {M.}~\bibnamefont {Ueda}},\ }\href {\doibase
  10.1103/PhysRevX.8.031079} {\bibfield  {journal} {\bibinfo  {journal}
  {Physical Review X}\ }\textbf {\bibinfo {volume} {8}},\ \bibinfo {pages}
  {031079} (\bibinfo {year} {2018})},\ \Eprint
  {http://arxiv.org/abs/1802.07964} {arXiv:1802.07964} \BibitemShut {NoStop}%
\bibitem [{\citenamefont {Lieu}(2018{\natexlab{a}})}]{lieu2018topological}%
  \BibitemOpen
  \bibfield  {author} {\bibinfo {author} {\bibfnamefont {S.}~\bibnamefont
  {Lieu}},\ }\href {\doibase 10.1103/PhysRevB.98.115135} {\bibfield  {journal}
  {\bibinfo  {journal} {Physical Review B}\ }\textbf {\bibinfo {volume} {98}},\
  \bibinfo {pages} {115135} (\bibinfo {year} {2018}{\natexlab{a}})}\BibitemShut
  {NoStop}%
\bibitem [{\citenamefont {Lieu}(2018{\natexlab{b}})}]{lieu2018topological1}%
  \BibitemOpen
  \bibfield  {author} {\bibinfo {author} {\bibfnamefont {S.}~\bibnamefont
  {Lieu}},\ }\href {\doibase 10.1103/PhysRevB.97.045106} {\bibfield  {journal}
  {\bibinfo  {journal} {Physical Review B}\ }\textbf {\bibinfo {volume} {97}},\
  \bibinfo {pages} {045106} (\bibinfo {year} {2018}{\natexlab{b}})}\BibitemShut
  {NoStop}%
\bibitem [{\citenamefont {Kunst}\ \emph {et~al.}(2018)\citenamefont {Kunst},
  \citenamefont {Edvardsson}, \citenamefont {Budich},\ and\ \citenamefont
  {Bergholtz}}]{kunst2018biorthogonal}%
  \BibitemOpen
  \bibfield  {author} {\bibinfo {author} {\bibfnamefont {F.~K.}\ \bibnamefont
  {Kunst}}, \bibinfo {author} {\bibfnamefont {E.}~\bibnamefont {Edvardsson}},
  \bibinfo {author} {\bibfnamefont {J.~C.}\ \bibnamefont {Budich}}, \ and\
  \bibinfo {author} {\bibfnamefont {E.~J.}\ \bibnamefont {Bergholtz}},\ }\href
  {\doibase 10.1103/PhysRevLett.121.026808} {\bibfield  {journal} {\bibinfo
  {journal} {Physical Review Letters}\ }\textbf {\bibinfo {volume} {121}},\
  \bibinfo {pages} {026808} (\bibinfo {year} {2018})}\BibitemShut {NoStop}%
\bibitem [{\citenamefont {Kawabata}\ \emph {et~al.}(2018)\citenamefont
  {Kawabata}, \citenamefont {Shiozaki},\ and\ \citenamefont
  {Ueda}}]{kawabata2018non}%
  \BibitemOpen
  \bibfield  {author} {\bibinfo {author} {\bibfnamefont {K.}~\bibnamefont
  {Kawabata}}, \bibinfo {author} {\bibfnamefont {K.}~\bibnamefont {Shiozaki}},
  \ and\ \bibinfo {author} {\bibfnamefont {M.}~\bibnamefont {Ueda}},\ }\href
  {http://arxiv.org/abs/1805.09632} {\  (\bibinfo {year} {2018})},\ \Eprint
  {http://arxiv.org/abs/1805.09632} {arXiv:1805.09632} \BibitemShut {NoStop}%
\bibitem [{\citenamefont {Yao}\ \emph {et~al.}(2018)\citenamefont {Yao},
  \citenamefont {Song},\ and\ \citenamefont {Wang}}]{yao2018non}%
  \BibitemOpen
  \bibfield  {author} {\bibinfo {author} {\bibfnamefont {S.}~\bibnamefont
  {Yao}}, \bibinfo {author} {\bibfnamefont {F.}~\bibnamefont {Song}}, \ and\
  \bibinfo {author} {\bibfnamefont {Z.}~\bibnamefont {Wang}},\ }\href {\doibase
  10.1103/PhysRevLett.121.136802} {\bibfield  {journal} {\bibinfo  {journal}
  {Physical Review Letters}\ }\textbf {\bibinfo {volume} {121}},\ \bibinfo
  {pages} {136802} (\bibinfo {year} {2018})},\ \Eprint
  {http://arxiv.org/abs/1804.04672} {arXiv:1804.04672} \BibitemShut {NoStop}%
\bibitem [{\citenamefont {Yao}\ and\ \citenamefont {Wang}(2018)}]{yao2018edge}%
  \BibitemOpen
  \bibfield  {author} {\bibinfo {author} {\bibfnamefont {S.}~\bibnamefont
  {Yao}}\ and\ \bibinfo {author} {\bibfnamefont {Z.}~\bibnamefont {Wang}},\
  }\href {\doibase 10.1103/PhysRevLett.121.086803} {\bibfield  {journal}
  {\bibinfo  {journal} {Physical Review Letters}\ }\textbf {\bibinfo {volume}
  {121}},\ \bibinfo {pages} {086803} (\bibinfo {year} {2018})},\ \Eprint
  {http://arxiv.org/abs/1803.01876} {arXiv:1803.01876} \BibitemShut {NoStop}%
\bibitem [{\citenamefont {Xiong}(2018)}]{xiong2018why}%
  \BibitemOpen
  \bibfield  {author} {\bibinfo {author} {\bibfnamefont {Y.}~\bibnamefont
  {Xiong}},\ }\href {\doibase 10.1088/2399-6528/aab64a} {\bibfield  {journal}
  {\bibinfo  {journal} {Journal of Physics Communications}\ }\textbf {\bibinfo
  {volume} {2}},\ \bibinfo {pages} {035043} (\bibinfo {year} {2018})},\ \Eprint
  {http://arxiv.org/abs/1705.06039} {arXiv:1705.06039} \BibitemShut {NoStop}%
\bibitem [{\citenamefont {Lee}\ and\ \citenamefont
  {Thomale}(2018)}]{lee2018anatomy}%
  \BibitemOpen
  \bibfield  {author} {\bibinfo {author} {\bibfnamefont {C.~H.}\ \bibnamefont
  {Lee}}\ and\ \bibinfo {author} {\bibfnamefont {R.}~\bibnamefont {Thomale}},\
  }\href {http://arxiv.org/abs/1809.02125} {\bibfield  {journal} {\bibinfo
  {journal} {arXiv}\ } (\bibinfo {year} {2018})},\ \Eprint
  {http://arxiv.org/abs/1809.02125} {arXiv:1809.02125} \BibitemShut {NoStop}%
\bibitem [{\citenamefont {Carlstr{\"{o}}m}\ and\ \citenamefont
  {Bergholtz}(2018)}]{carlstrom2018exceptional}%
  \BibitemOpen
  \bibfield  {author} {\bibinfo {author} {\bibfnamefont {J.}~\bibnamefont
  {Carlstr{\"{o}}m}}\ and\ \bibinfo {author} {\bibfnamefont {E.~J.}\
  \bibnamefont {Bergholtz}},\ }\href {http://arxiv.org/abs/1807.03330}
  {\bibfield  {journal} {\bibinfo  {journal} {arXiv}\ } (\bibinfo {year}
  {2018})},\ \Eprint {http://arxiv.org/abs/1807.03330} {arXiv:1807.03330}
  \BibitemShut {NoStop}%
\bibitem [{\citenamefont {Dembowski}\ \emph {et~al.}(2001)\citenamefont
  {Dembowski}, \citenamefont {Gr{\"{a}}f}, \citenamefont {Harney},
  \citenamefont {Heine}, \citenamefont {Heiss}, \citenamefont {Rehfeld},\ and\
  \citenamefont {Richter}}]{dembowski2001experimental}%
  \BibitemOpen
  \bibfield  {author} {\bibinfo {author} {\bibfnamefont {C.}~\bibnamefont
  {Dembowski}}, \bibinfo {author} {\bibfnamefont {H.-D.}\ \bibnamefont
  {Gr{\"{a}}f}}, \bibinfo {author} {\bibfnamefont {H.~L.}\ \bibnamefont
  {Harney}}, \bibinfo {author} {\bibfnamefont {A.}~\bibnamefont {Heine}},
  \bibinfo {author} {\bibfnamefont {W.~D.}\ \bibnamefont {Heiss}}, \bibinfo
  {author} {\bibfnamefont {H.}~\bibnamefont {Rehfeld}}, \ and\ \bibinfo
  {author} {\bibfnamefont {A.}~\bibnamefont {Richter}},\ }\href {\doibase
  10.1103/PhysRevLett.86.787} {\bibfield  {journal} {\bibinfo  {journal}
  {Physical Review Letters}\ }\textbf {\bibinfo {volume} {86}},\ \bibinfo
  {pages} {787} (\bibinfo {year} {2001})}\BibitemShut {NoStop}%
\bibitem [{\citenamefont {Poli}\ \emph {et~al.}(2015)\citenamefont {Poli},
  \citenamefont {Bellec}, \citenamefont {Kuhl}, \citenamefont {Mortessagne},\
  and\ \citenamefont {Schomerus}}]{poli2015selective}%
  \BibitemOpen
  \bibfield  {author} {\bibinfo {author} {\bibfnamefont {C.}~\bibnamefont
  {Poli}}, \bibinfo {author} {\bibfnamefont {M.}~\bibnamefont {Bellec}},
  \bibinfo {author} {\bibfnamefont {U.}~\bibnamefont {Kuhl}}, \bibinfo {author}
  {\bibfnamefont {F.}~\bibnamefont {Mortessagne}}, \ and\ \bibinfo {author}
  {\bibfnamefont {H.}~\bibnamefont {Schomerus}},\ }\href {\doibase
  10.1038/ncomms7710} {\bibfield  {journal} {\bibinfo  {journal} {Nature
  Communications}\ }\textbf {\bibinfo {volume} {6}},\ \bibinfo {pages} {6710}
  (\bibinfo {year} {2015})}\BibitemShut {NoStop}%
\bibitem [{\citenamefont {Zeuner}\ \emph {et~al.}(2015)\citenamefont {Zeuner},
  \citenamefont {Rechtsman}, \citenamefont {Plotnik}, \citenamefont {Lumer},
  \citenamefont {Nolte}, \citenamefont {Rudner}, \citenamefont {Segev},\ and\
  \citenamefont {Szameit}}]{zeuner2015observation}%
  \BibitemOpen
  \bibfield  {author} {\bibinfo {author} {\bibfnamefont {J.~M.}\ \bibnamefont
  {Zeuner}}, \bibinfo {author} {\bibfnamefont {M.~C.}\ \bibnamefont
  {Rechtsman}}, \bibinfo {author} {\bibfnamefont {Y.}~\bibnamefont {Plotnik}},
  \bibinfo {author} {\bibfnamefont {Y.}~\bibnamefont {Lumer}}, \bibinfo
  {author} {\bibfnamefont {S.}~\bibnamefont {Nolte}}, \bibinfo {author}
  {\bibfnamefont {M.~S.}\ \bibnamefont {Rudner}}, \bibinfo {author}
  {\bibfnamefont {M.}~\bibnamefont {Segev}}, \ and\ \bibinfo {author}
  {\bibfnamefont {A.}~\bibnamefont {Szameit}},\ }\href {\doibase
  10.1103/PhysRevLett.115.040402} {\bibfield  {journal} {\bibinfo  {journal}
  {Physical Review Letters}\ }\textbf {\bibinfo {volume} {115}},\ \bibinfo
  {pages} {040402} (\bibinfo {year} {2015})}\BibitemShut {NoStop}%
\bibitem [{\citenamefont {Weimann}\ \emph {et~al.}(2016)\citenamefont
  {Weimann}, \citenamefont {Kremer}, \citenamefont {Plotnik}, \citenamefont
  {Lumer}, \citenamefont {Nolte}, \citenamefont {Makris}, \citenamefont
  {Segev}, \citenamefont {Rechtsman},\ and\ \citenamefont
  {Szameit}}]{weimann2016topologically}%
  \BibitemOpen
  \bibfield  {author} {\bibinfo {author} {\bibfnamefont {S.}~\bibnamefont
  {Weimann}}, \bibinfo {author} {\bibfnamefont {M.}~\bibnamefont {Kremer}},
  \bibinfo {author} {\bibfnamefont {Y.}~\bibnamefont {Plotnik}}, \bibinfo
  {author} {\bibfnamefont {Y.}~\bibnamefont {Lumer}}, \bibinfo {author}
  {\bibfnamefont {S.}~\bibnamefont {Nolte}}, \bibinfo {author} {\bibfnamefont
  {K.~G.}\ \bibnamefont {Makris}}, \bibinfo {author} {\bibfnamefont
  {M.}~\bibnamefont {Segev}}, \bibinfo {author} {\bibfnamefont
  {M.}~\bibnamefont {Rechtsman}}, \ and\ \bibinfo {author} {\bibfnamefont
  {A.}~\bibnamefont {Szameit}},\ }\href {\doibase 10.1038/nmat4811} {\bibfield
  {journal} {\bibinfo  {journal} {Nature Materials}\ }\textbf {\bibinfo
  {volume} {16}},\ \bibinfo {pages} {433} (\bibinfo {year} {2016})}\BibitemShut
  {NoStop}%
\bibitem [{\citenamefont {Zhou}\ \emph {et~al.}(2018)\citenamefont {Zhou},
  \citenamefont {Peng}, \citenamefont {Yoon}, \citenamefont {Hsu},
  \citenamefont {Nelson}, \citenamefont {Fu}, \citenamefont {Joannopoulos},
  \citenamefont {Solja{\v{c}}i{\'{c}}},\ and\ \citenamefont
  {Zhen}}]{zhou2018observation}%
  \BibitemOpen
  \bibfield  {author} {\bibinfo {author} {\bibfnamefont {H.}~\bibnamefont
  {Zhou}}, \bibinfo {author} {\bibfnamefont {C.}~\bibnamefont {Peng}}, \bibinfo
  {author} {\bibfnamefont {Y.}~\bibnamefont {Yoon}}, \bibinfo {author}
  {\bibfnamefont {C.~W.}\ \bibnamefont {Hsu}}, \bibinfo {author} {\bibfnamefont
  {K.~A.}\ \bibnamefont {Nelson}}, \bibinfo {author} {\bibfnamefont
  {L.}~\bibnamefont {Fu}}, \bibinfo {author} {\bibfnamefont {J.~D.}\
  \bibnamefont {Joannopoulos}}, \bibinfo {author} {\bibfnamefont
  {M.}~\bibnamefont {Solja{\v{c}}i{\'{c}}}}, \ and\ \bibinfo {author}
  {\bibfnamefont {B.}~\bibnamefont {Zhen}},\ }\href {\doibase
  10.1126/science.aap9859} {\bibfield  {journal} {\bibinfo  {journal}
  {Science}\ }\textbf {\bibinfo {volume} {359}},\ \bibinfo {pages} {1009}
  (\bibinfo {year} {2018})}\BibitemShut {NoStop}%
\bibitem [{\citenamefont {Cerjan}\ \emph
  {et~al.}(2018{\natexlab{a}})\citenamefont {Cerjan}, \citenamefont {Huang},
  \citenamefont {Chen}, \citenamefont {Chong},\ and\ \citenamefont
  {Rechtsman}}]{cerjan2018experimental}%
  \BibitemOpen
  \bibfield  {author} {\bibinfo {author} {\bibfnamefont {A.}~\bibnamefont
  {Cerjan}}, \bibinfo {author} {\bibfnamefont {S.}~\bibnamefont {Huang}},
  \bibinfo {author} {\bibfnamefont {K.~P.}\ \bibnamefont {Chen}}, \bibinfo
  {author} {\bibfnamefont {Y.}~\bibnamefont {Chong}}, \ and\ \bibinfo {author}
  {\bibfnamefont {M.~C.}\ \bibnamefont {Rechtsman}},\ }\href
  {https://arxiv.org/abs/1808.09541} {\bibfield  {journal} {\bibinfo  {journal}
  {arXiv}\ } (\bibinfo {year} {2018}{\natexlab{a}})},\ \Eprint
  {http://arxiv.org/abs/1808.09541} {arXiv:1808.09541} \BibitemShut {NoStop}%
\bibitem [{\citenamefont {Lin}\ \emph {et~al.}(2011)\citenamefont {Lin},
  \citenamefont {Ramezani}, \citenamefont {Eichelkraut}, \citenamefont
  {Kottos}, \citenamefont {Cao},\ and\ \citenamefont
  {Christodoulides}}]{lin2011unidirectional}%
  \BibitemOpen
  \bibfield  {author} {\bibinfo {author} {\bibfnamefont {Z.}~\bibnamefont
  {Lin}}, \bibinfo {author} {\bibfnamefont {H.}~\bibnamefont {Ramezani}},
  \bibinfo {author} {\bibfnamefont {T.}~\bibnamefont {Eichelkraut}}, \bibinfo
  {author} {\bibfnamefont {T.}~\bibnamefont {Kottos}}, \bibinfo {author}
  {\bibfnamefont {H.}~\bibnamefont {Cao}}, \ and\ \bibinfo {author}
  {\bibfnamefont {D.~N.}\ \bibnamefont {Christodoulides}},\ }\href {\doibase
  10.1103/PhysRevLett.106.213901} {\bibfield  {journal} {\bibinfo  {journal}
  {Physical Review Letters}\ }\textbf {\bibinfo {volume} {106}},\ \bibinfo
  {pages} {213901} (\bibinfo {year} {2011})}\BibitemShut {NoStop}%
\bibitem [{\citenamefont {Feng}\ \emph {et~al.}(2013)\citenamefont {Feng},
  \citenamefont {Xu}, \citenamefont {Fegadolli}, \citenamefont {Lu},
  \citenamefont {Oliveira}, \citenamefont {Almeida}, \citenamefont {Chen},\
  and\ \citenamefont {Scherer}}]{feng2013experimental}%
  \BibitemOpen
  \bibfield  {author} {\bibinfo {author} {\bibfnamefont {L.}~\bibnamefont
  {Feng}}, \bibinfo {author} {\bibfnamefont {Y.-L.}\ \bibnamefont {Xu}},
  \bibinfo {author} {\bibfnamefont {W.~S.}\ \bibnamefont {Fegadolli}}, \bibinfo
  {author} {\bibfnamefont {M.-H.}\ \bibnamefont {Lu}}, \bibinfo {author}
  {\bibfnamefont {J.~E.~B.}\ \bibnamefont {Oliveira}}, \bibinfo {author}
  {\bibfnamefont {V.~R.}\ \bibnamefont {Almeida}}, \bibinfo {author}
  {\bibfnamefont {Y.-F.}\ \bibnamefont {Chen}}, \ and\ \bibinfo {author}
  {\bibfnamefont {A.}~\bibnamefont {Scherer}},\ }\href@noop {} {\bibfield
  {journal} {\bibinfo  {journal} {Nature Materials}\ }\textbf {\bibinfo
  {volume} {12}},\ \bibinfo {pages} {108} (\bibinfo {year} {2013})}\BibitemShut
  {NoStop}%
\bibitem [{\citenamefont {Chen}\ \emph {et~al.}(2017)\citenamefont {Chen},
  \citenamefont {{\"{O}}zdemir}, \citenamefont {Zhao}, \citenamefont
  {Wiersig},\ and\ \citenamefont {Yang}}]{chen2017exceptional}%
  \BibitemOpen
  \bibfield  {author} {\bibinfo {author} {\bibfnamefont {W.}~\bibnamefont
  {Chen}}, \bibinfo {author} {\bibfnamefont {Å.~K.}\ \bibnamefont
  {{\"{O}}zdemir}}, \bibinfo {author} {\bibfnamefont {G.}~\bibnamefont {Zhao}},
  \bibinfo {author} {\bibfnamefont {J.}~\bibnamefont {Wiersig}}, \ and\
  \bibinfo {author} {\bibfnamefont {L.}~\bibnamefont {Yang}},\ }\href {\doibase
  10.1038/nature23281} {\bibfield  {journal} {\bibinfo  {journal} {Nature}\
  }\textbf {\bibinfo {volume} {548}},\ \bibinfo {pages} {192} (\bibinfo {year}
  {2017})}\BibitemShut {NoStop}%
\bibitem [{\citenamefont {Hodaei}\ \emph {et~al.}(2017)\citenamefont {Hodaei},
  \citenamefont {Hassan}, \citenamefont {Wittek}, \citenamefont
  {Garcia-Gracia}, \citenamefont {El-Ganainy}, \citenamefont
  {Christodoulides},\ and\ \citenamefont {Khajavikhan}}]{hodaei2017enhanced}%
  \BibitemOpen
  \bibfield  {author} {\bibinfo {author} {\bibfnamefont {H.}~\bibnamefont
  {Hodaei}}, \bibinfo {author} {\bibfnamefont {A.~U.}\ \bibnamefont {Hassan}},
  \bibinfo {author} {\bibfnamefont {S.}~\bibnamefont {Wittek}}, \bibinfo
  {author} {\bibfnamefont {H.}~\bibnamefont {Garcia-Gracia}}, \bibinfo {author}
  {\bibfnamefont {R.}~\bibnamefont {El-Ganainy}}, \bibinfo {author}
  {\bibfnamefont {D.~N.}\ \bibnamefont {Christodoulides}}, \ and\ \bibinfo
  {author} {\bibfnamefont {M.}~\bibnamefont {Khajavikhan}},\ }\href {\doibase
  10.1038/nature23280} {\bibfield  {journal} {\bibinfo  {journal} {Nature}\
  }\textbf {\bibinfo {volume} {548}},\ \bibinfo {pages} {187} (\bibinfo {year}
  {2017})}\BibitemShut {NoStop}%
\bibitem [{\citenamefont {Zhang}\ \emph {et~al.}(2018)\citenamefont {Zhang},
  \citenamefont {Sweeney}, \citenamefont {Hsu}, \citenamefont {Yang},
  \citenamefont {Stone},\ and\ \citenamefont {Jiang}}]{zhang2018quantum}%
  \BibitemOpen
  \bibfield  {author} {\bibinfo {author} {\bibfnamefont {M.}~\bibnamefont
  {Zhang}}, \bibinfo {author} {\bibfnamefont {W.}~\bibnamefont {Sweeney}},
  \bibinfo {author} {\bibfnamefont {C.~W.}\ \bibnamefont {Hsu}}, \bibinfo
  {author} {\bibfnamefont {L.}~\bibnamefont {Yang}}, \bibinfo {author}
  {\bibfnamefont {A.~D.}\ \bibnamefont {Stone}}, \ and\ \bibinfo {author}
  {\bibfnamefont {L.}~\bibnamefont {Jiang}},\ }\href
  {http://arxiv.org/abs/1805.12001} {\bibfield  {journal} {\bibinfo  {journal}
  {arXiv}\ } (\bibinfo {year} {2018})},\ \Eprint
  {http://arxiv.org/abs/1805.12001} {arXiv:1805.12001} \BibitemShut {NoStop}%
\bibitem [{\citenamefont {Lau}\ and\ \citenamefont {Clerk}(2018)}]{lau2018non}%
  \BibitemOpen
  \bibfield  {author} {\bibinfo {author} {\bibfnamefont {H.-K.}\ \bibnamefont
  {Lau}}\ and\ \bibinfo {author} {\bibfnamefont {A.~A.}\ \bibnamefont
  {Clerk}},\ }\href {http://arxiv.org/abs/1805.11760} {\bibfield  {journal}
  {\bibinfo  {journal} {arXiv}\ } (\bibinfo {year} {2018})},\ \Eprint
  {http://arxiv.org/abs/1805.11760} {arXiv:1805.11760} \BibitemShut {NoStop}%
\bibitem [{\citenamefont {Hodaei}\ \emph {et~al.}(2014)\citenamefont {Hodaei},
  \citenamefont {Miri}, \citenamefont {Heinrich}, \citenamefont
  {Christodoulides},\ and\ \citenamefont {Khajavikhan}}]{hodaei2014parity}%
  \BibitemOpen
  \bibfield  {author} {\bibinfo {author} {\bibfnamefont {H.}~\bibnamefont
  {Hodaei}}, \bibinfo {author} {\bibfnamefont {M.-A.}\ \bibnamefont {Miri}},
  \bibinfo {author} {\bibfnamefont {M.}~\bibnamefont {Heinrich}}, \bibinfo
  {author} {\bibfnamefont {D.~N.}\ \bibnamefont {Christodoulides}}, \ and\
  \bibinfo {author} {\bibfnamefont {M.}~\bibnamefont {Khajavikhan}},\ }\href
  {\doibase 10.1126/science.1258480} {\bibfield  {journal} {\bibinfo  {journal}
  {Science}\ }\textbf {\bibinfo {volume} {346}},\ \bibinfo {pages} {975}
  (\bibinfo {year} {2014})}\BibitemShut {NoStop}%
\bibitem [{\citenamefont {Feng}\ \emph {et~al.}(2014)\citenamefont {Feng},
  \citenamefont {Wong}, \citenamefont {Ma}, \citenamefont {Wang},\ and\
  \citenamefont {Zhang}}]{feng2014single}%
  \BibitemOpen
  \bibfield  {author} {\bibinfo {author} {\bibfnamefont {L.}~\bibnamefont
  {Feng}}, \bibinfo {author} {\bibfnamefont {Z.~J.}\ \bibnamefont {Wong}},
  \bibinfo {author} {\bibfnamefont {R.-M.}\ \bibnamefont {Ma}}, \bibinfo
  {author} {\bibfnamefont {Y.}~\bibnamefont {Wang}}, \ and\ \bibinfo {author}
  {\bibfnamefont {X.}~\bibnamefont {Zhang}},\ }\href {\doibase
  10.1126/science.1258479} {\bibfield  {journal} {\bibinfo  {journal}
  {Science}\ }\textbf {\bibinfo {volume} {346}},\ \bibinfo {pages} {972}
  (\bibinfo {year} {2014})},\ \Eprint {http://arxiv.org/abs/0504102}
  {arXiv:0504102 [arXiv:physics]} \BibitemShut {NoStop}%
\bibitem [{\citenamefont {Peng}\ \emph {et~al.}(2014)\citenamefont {Peng},
  \citenamefont {Ozdemir}, \citenamefont {Rotter}, \citenamefont {Yilmaz},
  \citenamefont {Liertzer}, \citenamefont {Monifi}, \citenamefont {Bender},
  \citenamefont {Nori},\ and\ \citenamefont {Yang}}]{peng2014loss}%
  \BibitemOpen
  \bibfield  {author} {\bibinfo {author} {\bibfnamefont {B.}~\bibnamefont
  {Peng}}, \bibinfo {author} {\bibfnamefont {S.~K.}\ \bibnamefont {Ozdemir}},
  \bibinfo {author} {\bibfnamefont {S.}~\bibnamefont {Rotter}}, \bibinfo
  {author} {\bibfnamefont {H.}~\bibnamefont {Yilmaz}}, \bibinfo {author}
  {\bibfnamefont {M.}~\bibnamefont {Liertzer}}, \bibinfo {author}
  {\bibfnamefont {F.}~\bibnamefont {Monifi}}, \bibinfo {author} {\bibfnamefont
  {C.~M.}\ \bibnamefont {Bender}}, \bibinfo {author} {\bibfnamefont
  {F.}~\bibnamefont {Nori}}, \ and\ \bibinfo {author} {\bibfnamefont
  {L.}~\bibnamefont {Yang}},\ }\href {\doibase 10.1126/science.1258004}
  {\bibfield  {journal} {\bibinfo  {journal} {Science}\ }\textbf {\bibinfo
  {volume} {346}},\ \bibinfo {pages} {328} (\bibinfo {year} {2014})},\ \Eprint
  {http://arxiv.org/abs/1410.7474} {arXiv:1410.7474} \BibitemShut {NoStop}%
\bibitem [{\citenamefont {Harari}\ \emph {et~al.}(2018)\citenamefont {Harari},
  \citenamefont {Bandres}, \citenamefont {Lumer}, \citenamefont {Rechtsman},
  \citenamefont {Chong}, \citenamefont {Khajavikhan}, \citenamefont
  {Christodoulides},\ and\ \citenamefont {Segev}}]{harari2018topological}%
  \BibitemOpen
  \bibfield  {author} {\bibinfo {author} {\bibfnamefont {G.}~\bibnamefont
  {Harari}}, \bibinfo {author} {\bibfnamefont {M.~A.}\ \bibnamefont {Bandres}},
  \bibinfo {author} {\bibfnamefont {Y.}~\bibnamefont {Lumer}}, \bibinfo
  {author} {\bibfnamefont {M.~C.}\ \bibnamefont {Rechtsman}}, \bibinfo {author}
  {\bibfnamefont {Y.~D.}\ \bibnamefont {Chong}}, \bibinfo {author}
  {\bibfnamefont {M.}~\bibnamefont {Khajavikhan}}, \bibinfo {author}
  {\bibfnamefont {D.~N.}\ \bibnamefont {Christodoulides}}, \ and\ \bibinfo
  {author} {\bibfnamefont {M.}~\bibnamefont {Segev}},\ }\href {\doibase
  10.1126/science.aar4003} {\bibfield  {journal} {\bibinfo  {journal}
  {Science}\ }\textbf {\bibinfo {volume} {359}},\ \bibinfo {pages} {eaar4003}
  (\bibinfo {year} {2018})}\BibitemShut {NoStop}%
\bibitem [{\citenamefont {Bandres}\ \emph {et~al.}(2018)\citenamefont
  {Bandres}, \citenamefont {Wittek}, \citenamefont {Harari}, \citenamefont
  {Parto}, \citenamefont {Ren}, \citenamefont {Segev}, \citenamefont
  {Christodoulides},\ and\ \citenamefont
  {Khajavikhan}}]{bandres2018topological}%
  \BibitemOpen
  \bibfield  {author} {\bibinfo {author} {\bibfnamefont {M.~A.}\ \bibnamefont
  {Bandres}}, \bibinfo {author} {\bibfnamefont {S.}~\bibnamefont {Wittek}},
  \bibinfo {author} {\bibfnamefont {G.}~\bibnamefont {Harari}}, \bibinfo
  {author} {\bibfnamefont {M.}~\bibnamefont {Parto}}, \bibinfo {author}
  {\bibfnamefont {J.}~\bibnamefont {Ren}}, \bibinfo {author} {\bibfnamefont
  {M.}~\bibnamefont {Segev}}, \bibinfo {author} {\bibfnamefont {D.~N.}\
  \bibnamefont {Christodoulides}}, \ and\ \bibinfo {author} {\bibfnamefont
  {M.}~\bibnamefont {Khajavikhan}},\ }\href {\doibase 10.1126/science.aar4005}
  {\bibfield  {journal} {\bibinfo  {journal} {Science}\ }\textbf {\bibinfo
  {volume} {359}},\ \bibinfo {pages} {eaar4005} (\bibinfo {year}
  {2018})}\BibitemShut {NoStop}%
\bibitem [{\citenamefont {Xu}\ \emph {et~al.}(2016)\citenamefont {Xu},
  \citenamefont {Mason}, \citenamefont {Jiang},\ and\ \citenamefont
  {Harris}}]{xu2016topological}%
  \BibitemOpen
  \bibfield  {author} {\bibinfo {author} {\bibfnamefont {H.}~\bibnamefont
  {Xu}}, \bibinfo {author} {\bibfnamefont {D.}~\bibnamefont {Mason}}, \bibinfo
  {author} {\bibfnamefont {L.}~\bibnamefont {Jiang}}, \ and\ \bibinfo {author}
  {\bibfnamefont {J.~G.~E.}\ \bibnamefont {Harris}},\ }\href
  {http://dx.doi.org/10.1038/nature18604 http://10.0.4.14/nature18604
  http://arxiv.org/abs/1602.06881} {\bibfield  {journal} {\bibinfo  {journal}
  {Nature}\ }\textbf {\bibinfo {volume} {537}},\ \bibinfo {pages} {80}
  (\bibinfo {year} {2016})},\ \Eprint {http://arxiv.org/abs/1602.06881}
  {arXiv:1602.06881} \BibitemShut {NoStop}%
\bibitem [{\citenamefont {Doppler}\ \emph {et~al.}(2016)\citenamefont
  {Doppler}, \citenamefont {Mailybaev}, \citenamefont {B{\"{o}}hm},
  \citenamefont {Kuhl}, \citenamefont {Girschik}, \citenamefont {Libisch},
  \citenamefont {Milburn}, \citenamefont {Rabl}, \citenamefont {Moiseyev},\
  and\ \citenamefont {Rotter}}]{doppler2016dynamically}%
  \BibitemOpen
  \bibfield  {author} {\bibinfo {author} {\bibfnamefont {J.}~\bibnamefont
  {Doppler}}, \bibinfo {author} {\bibfnamefont {A.~A.}\ \bibnamefont
  {Mailybaev}}, \bibinfo {author} {\bibfnamefont {J.}~\bibnamefont
  {B{\"{o}}hm}}, \bibinfo {author} {\bibfnamefont {U.}~\bibnamefont {Kuhl}},
  \bibinfo {author} {\bibfnamefont {A.}~\bibnamefont {Girschik}}, \bibinfo
  {author} {\bibfnamefont {F.}~\bibnamefont {Libisch}}, \bibinfo {author}
  {\bibfnamefont {T.~J.}\ \bibnamefont {Milburn}}, \bibinfo {author}
  {\bibfnamefont {P.}~\bibnamefont {Rabl}}, \bibinfo {author} {\bibfnamefont
  {N.}~\bibnamefont {Moiseyev}}, \ and\ \bibinfo {author} {\bibfnamefont
  {S.}~\bibnamefont {Rotter}},\ }\href {http://dx.doi.org/10.1038/nature18605
  http://10.0.4.14/nature18605
  http://www.nature.com/nature/journal/v537/n7618/abs/nature18605.html{\#}supplementary-information}
  {\bibfield  {journal} {\bibinfo  {journal} {Nature}\ }\textbf {\bibinfo
  {volume} {537}},\ \bibinfo {pages} {76} (\bibinfo {year} {2016})}\BibitemShut
  {NoStop}%
\bibitem [{\citenamefont {Rotter}(2009)}]{rotter2009non}%
  \BibitemOpen
  \bibfield  {author} {\bibinfo {author} {\bibfnamefont {I.}~\bibnamefont
  {Rotter}},\ }\href {http://stacks.iop.org/1751-8121/42/i=15/a=153001}
  {\bibfield  {journal} {\bibinfo  {journal} {Journal of Physics A:
  Mathematical and General}\ }\textbf {\bibinfo {volume} {42}},\ \bibinfo
  {pages} {153001} (\bibinfo {year} {2009})}\BibitemShut {NoStop}%
\bibitem [{\citenamefont {Mailybaev}\ \emph {et~al.}(2005)\citenamefont
  {Mailybaev}, \citenamefont {Kirillov},\ and\ \citenamefont
  {Seyranian}}]{mailybaev2005geometric}%
  \BibitemOpen
  \bibfield  {author} {\bibinfo {author} {\bibfnamefont {A.~A.}\ \bibnamefont
  {Mailybaev}}, \bibinfo {author} {\bibfnamefont {O.~N.}\ \bibnamefont
  {Kirillov}}, \ and\ \bibinfo {author} {\bibfnamefont {A.~P.}\ \bibnamefont
  {Seyranian}},\ }\href {\doibase 10.1103/PhysRevA.72.014104} {\bibfield
  {journal} {\bibinfo  {journal} {Physical Review A}\ }\textbf {\bibinfo
  {volume} {72}},\ \bibinfo {pages} {014104} (\bibinfo {year} {2005})},\
  \Eprint {http://arxiv.org/abs/0501040} {arXiv:0501040 [quant-ph]}
  \BibitemShut {NoStop}%
\bibitem [{\citenamefont {Kozii}\ and\ \citenamefont
  {Fu}(2017)}]{kozii2017non}%
  \BibitemOpen
  \bibfield  {author} {\bibinfo {author} {\bibfnamefont {V.}~\bibnamefont
  {Kozii}}\ and\ \bibinfo {author} {\bibfnamefont {L.}~\bibnamefont {Fu}},\
  }\href {https://arxiv.org/abs/1708.05841 http://arxiv.org/abs/1708.05841}
  {\bibfield  {journal} {\bibinfo  {journal} {arXiv}\ } (\bibinfo {year}
  {2017})},\ \Eprint {http://arxiv.org/abs/1708.05841} {arXiv:1708.05841}
  \BibitemShut {NoStop}%
\bibitem [{\citenamefont {Liertzer}\ \emph {et~al.}(2012)\citenamefont
  {Liertzer}, \citenamefont {Ge}, \citenamefont {Cerjan}, \citenamefont
  {Stone}, \citenamefont {T{\"{u}}reci},\ and\ \citenamefont
  {Rotter}}]{liertzer2012pump}%
  \BibitemOpen
  \bibfield  {author} {\bibinfo {author} {\bibfnamefont {M.}~\bibnamefont
  {Liertzer}}, \bibinfo {author} {\bibfnamefont {L.}~\bibnamefont {Ge}},
  \bibinfo {author} {\bibfnamefont {A.}~\bibnamefont {Cerjan}}, \bibinfo
  {author} {\bibfnamefont {A.~D.}\ \bibnamefont {Stone}}, \bibinfo {author}
  {\bibfnamefont {H.~E.}\ \bibnamefont {T{\"{u}}reci}}, \ and\ \bibinfo
  {author} {\bibfnamefont {S.}~\bibnamefont {Rotter}},\ }\href {\doibase
  10.1103/PhysRevLett.108.173901} {\bibfield  {journal} {\bibinfo  {journal}
  {Physical Review Letters}\ }\textbf {\bibinfo {volume} {108}},\ \bibinfo
  {pages} {173901} (\bibinfo {year} {2012})},\ \Eprint
  {http://arxiv.org/abs/1109.0454} {arXiv:1109.0454} \BibitemShut {NoStop}%
\bibitem [{\citenamefont {Zhen}\ \emph {et~al.}(2015)\citenamefont {Zhen},
  \citenamefont {Hsu}, \citenamefont {Igarashi}, \citenamefont {Lu},
  \citenamefont {Kaminer}, \citenamefont {Pick}, \citenamefont {Chua},
  \citenamefont {Joannopoulos},\ and\ \citenamefont
  {Solja{\v{c}}i{\'{c}}}}]{zhen2015spawning}%
  \BibitemOpen
  \bibfield  {author} {\bibinfo {author} {\bibfnamefont {B.}~\bibnamefont
  {Zhen}}, \bibinfo {author} {\bibfnamefont {C.~W.}\ \bibnamefont {Hsu}},
  \bibinfo {author} {\bibfnamefont {Y.}~\bibnamefont {Igarashi}}, \bibinfo
  {author} {\bibfnamefont {L.}~\bibnamefont {Lu}}, \bibinfo {author}
  {\bibfnamefont {I.}~\bibnamefont {Kaminer}}, \bibinfo {author} {\bibfnamefont
  {A.}~\bibnamefont {Pick}}, \bibinfo {author} {\bibfnamefont {S.-L.}\
  \bibnamefont {Chua}}, \bibinfo {author} {\bibfnamefont {J.~D.}\ \bibnamefont
  {Joannopoulos}}, \ and\ \bibinfo {author} {\bibfnamefont {M.}~\bibnamefont
  {Solja{\v{c}}i{\'{c}}}},\ }\href {\doibase 10.1038/nature14889} {\bibfield
  {journal} {\bibinfo  {journal} {Nature}\ }\textbf {\bibinfo {volume} {525}},\
  \bibinfo {pages} {354} (\bibinfo {year} {2015})}\BibitemShut {NoStop}%
\bibitem [{\citenamefont {Cerjan}\ \emph {et~al.}(2016)\citenamefont {Cerjan},
  \citenamefont {Raman},\ and\ \citenamefont {Fan}}]{cerjan2016exceptional}%
  \BibitemOpen
  \bibfield  {author} {\bibinfo {author} {\bibfnamefont {A.}~\bibnamefont
  {Cerjan}}, \bibinfo {author} {\bibfnamefont {A.}~\bibnamefont {Raman}}, \
  and\ \bibinfo {author} {\bibfnamefont {S.}~\bibnamefont {Fan}},\ }\href
  {\doibase 10.1103/PhysRevLett.116.203902} {\bibfield  {journal} {\bibinfo
  {journal} {Physical Review Letters}\ }\textbf {\bibinfo {volume} {116}},\
  \bibinfo {pages} {203902} (\bibinfo {year} {2016})},\ \Eprint
  {http://arxiv.org/abs/1601.05489} {arXiv:1601.05489} \BibitemShut {NoStop}%
\bibitem [{\citenamefont {Xu}\ \emph {et~al.}(2017)\citenamefont {Xu},
  \citenamefont {Wang},\ and\ \citenamefont {Duan}}]{xu2017weyl}%
  \BibitemOpen
  \bibfield  {author} {\bibinfo {author} {\bibfnamefont {Y.}~\bibnamefont
  {Xu}}, \bibinfo {author} {\bibfnamefont {S.~T.}\ \bibnamefont {Wang}}, \ and\
  \bibinfo {author} {\bibfnamefont {L.~M.}\ \bibnamefont {Duan}},\ }\href
  {\doibase 10.1103/PhysRevLett.118.045701} {\bibfield  {journal} {\bibinfo
  {journal} {Physical Review Letters}\ }\textbf {\bibinfo {volume} {118}},\
  \bibinfo {pages} {045701} (\bibinfo {year} {2017})},\ \Eprint
  {http://arxiv.org/abs/1611.02239} {arXiv:1611.02239} \BibitemShut {NoStop}%
\bibitem [{\citenamefont {Cerjan}\ \emph
  {et~al.}(2018{\natexlab{b}})\citenamefont {Cerjan}, \citenamefont {Xiao},
  \citenamefont {Yuan},\ and\ \citenamefont {Fan}}]{cerjan2018effects}%
  \BibitemOpen
  \bibfield  {author} {\bibinfo {author} {\bibfnamefont {A.}~\bibnamefont
  {Cerjan}}, \bibinfo {author} {\bibfnamefont {M.}~\bibnamefont {Xiao}},
  \bibinfo {author} {\bibfnamefont {L.}~\bibnamefont {Yuan}}, \ and\ \bibinfo
  {author} {\bibfnamefont {S.}~\bibnamefont {Fan}},\ }\href {\doibase
  10.1103/PhysRevB.97.075128} {\bibfield  {journal} {\bibinfo  {journal}
  {Physical Review B}\ }\textbf {\bibinfo {volume} {97}} (\bibinfo {year}
  {2018}{\natexlab{b}}),\ 10.1103/PhysRevB.97.075128},\ \Eprint
  {http://arxiv.org/abs/1712.02444} {arXiv:1712.02444} \BibitemShut {NoStop}%
\bibitem [{\citenamefont {Fang}\ \emph {et~al.}(2016)\citenamefont {Fang},
  \citenamefont {Weng}, \citenamefont {Dai},\ and\ \citenamefont
  {Fang}}]{fang2016topological}%
  \BibitemOpen
  \bibfield  {author} {\bibinfo {author} {\bibfnamefont {C.}~\bibnamefont
  {Fang}}, \bibinfo {author} {\bibfnamefont {H.}~\bibnamefont {Weng}}, \bibinfo
  {author} {\bibfnamefont {X.}~\bibnamefont {Dai}}, \ and\ \bibinfo {author}
  {\bibfnamefont {Z.}~\bibnamefont {Fang}},\ }\href {\doibase
  10.1088/1674-1056/25/11/117106} {\bibfield  {journal} {\bibinfo  {journal}
  {Chinese Physics B}\ }\textbf {\bibinfo {volume} {25}},\ \bibinfo {pages}
  {117106} (\bibinfo {year} {2016})}\BibitemShut {NoStop}%
\bibitem [{\citenamefont {Chiu}\ \emph {et~al.}(2016)\citenamefont {Chiu},
  \citenamefont {Teo}, \citenamefont {Schnyder},\ and\ \citenamefont
  {Ryu}}]{chiu2016classification}%
  \BibitemOpen
  \bibfield  {author} {\bibinfo {author} {\bibfnamefont {C.-K.}\ \bibnamefont
  {Chiu}}, \bibinfo {author} {\bibfnamefont {J.~C.}\ \bibnamefont {Teo}},
  \bibinfo {author} {\bibfnamefont {A.~P.}\ \bibnamefont {Schnyder}}, \ and\
  \bibinfo {author} {\bibfnamefont {S.}~\bibnamefont {Ryu}},\ }\href {\doibase
  10.1103/RevModPhys.88.035005} {\bibfield  {journal} {\bibinfo  {journal}
  {Reviews of Modern Physics}\ }\textbf {\bibinfo {volume} {88}},\ \bibinfo
  {pages} {035005} (\bibinfo {year} {2016})}\BibitemShut {NoStop}%
\bibitem [{\citenamefont {Armitage}\ \emph {et~al.}(2018)\citenamefont
  {Armitage}, \citenamefont {Mele},\ and\ \citenamefont
  {Vishwanath}}]{armitage2018weyl}%
  \BibitemOpen
  \bibfield  {author} {\bibinfo {author} {\bibfnamefont {N.~P.}\ \bibnamefont
  {Armitage}}, \bibinfo {author} {\bibfnamefont {E.~J.}\ \bibnamefont {Mele}},
  \ and\ \bibinfo {author} {\bibfnamefont {A.}~\bibnamefont {Vishwanath}},\
  }\href {\doibase 10.1103/RevModPhys.90.015001} {\bibfield  {journal}
  {\bibinfo  {journal} {Reviews of Modern Physics}\ }\textbf {\bibinfo {volume}
  {90}},\ \bibinfo {pages} {015001} (\bibinfo {year} {2018})},\ \Eprint
  {http://arxiv.org/abs/1705.01111} {arXiv:1705.01111} \BibitemShut {NoStop}%
\bibitem [{\citenamefont {Burkov}\ \emph {et~al.}(2011)\citenamefont {Burkov},
  \citenamefont {Hook},\ and\ \citenamefont {Balents}}]{burkov2011topological}%
  \BibitemOpen
  \bibfield  {author} {\bibinfo {author} {\bibfnamefont {A.~A.}\ \bibnamefont
  {Burkov}}, \bibinfo {author} {\bibfnamefont {M.~D.}\ \bibnamefont {Hook}}, \
  and\ \bibinfo {author} {\bibfnamefont {L.}~\bibnamefont {Balents}},\ }\href
  {\doibase 10.1103/PhysRevB.84.235126} {\bibfield  {journal} {\bibinfo
  {journal} {Physical Review B}\ }\textbf {\bibinfo {volume} {84}},\ \bibinfo
  {pages} {235126} (\bibinfo {year} {2011})}\BibitemShut {NoStop}%
\bibitem [{\citenamefont {Fang}\ \emph {et~al.}(2015)\citenamefont {Fang},
  \citenamefont {Chen}, \citenamefont {Kee},\ and\ \citenamefont
  {Fu}}]{fang2015topological}%
  \BibitemOpen
  \bibfield  {author} {\bibinfo {author} {\bibfnamefont {C.}~\bibnamefont
  {Fang}}, \bibinfo {author} {\bibfnamefont {Y.}~\bibnamefont {Chen}}, \bibinfo
  {author} {\bibfnamefont {H.-Y.}\ \bibnamefont {Kee}}, \ and\ \bibinfo
  {author} {\bibfnamefont {L.}~\bibnamefont {Fu}},\ }\href {\doibase
  10.1103/PhysRevB.92.081201} {\bibfield  {journal} {\bibinfo  {journal}
  {Physical Review B}\ }\textbf {\bibinfo {volume} {92}},\ \bibinfo {pages}
  {081201} (\bibinfo {year} {2015})}\BibitemShut {NoStop}%
\bibitem [{\citenamefont {Zhang}\ \emph {et~al.}(2016)\citenamefont {Zhang},
  \citenamefont {Zhao}, \citenamefont {Liu}, \citenamefont {Xue}, \citenamefont
  {Zhu},\ and\ \citenamefont {Wang}}]{zhang2016quantum}%
  \BibitemOpen
  \bibfield  {author} {\bibinfo {author} {\bibfnamefont {D.-W.}\ \bibnamefont
  {Zhang}}, \bibinfo {author} {\bibfnamefont {Y.~X.}\ \bibnamefont {Zhao}},
  \bibinfo {author} {\bibfnamefont {R.-B.}\ \bibnamefont {Liu}}, \bibinfo
  {author} {\bibfnamefont {Z.-Y.}\ \bibnamefont {Xue}}, \bibinfo {author}
  {\bibfnamefont {S.-L.}\ \bibnamefont {Zhu}}, \ and\ \bibinfo {author}
  {\bibfnamefont {Z.~D.}\ \bibnamefont {Wang}},\ }\href {\doibase
  10.1103/PhysRevA.93.043617} {\bibfield  {journal} {\bibinfo  {journal}
  {Physical Review A}\ }\textbf {\bibinfo {volume} {93}},\ \bibinfo {pages}
  {043617} (\bibinfo {year} {2016})}\BibitemShut {NoStop}%
\bibitem [{\citenamefont {Bian}\ \emph {et~al.}(2016)\citenamefont {Bian},
  \citenamefont {Chang}, \citenamefont {Sankar}, \citenamefont {Xu},
  \citenamefont {Zheng}, \citenamefont {Neupert}, \citenamefont {Chiu},
  \citenamefont {Huang}, \citenamefont {Chang}, \citenamefont {Belopolski},
  \citenamefont {Sanchez}, \citenamefont {Neupane}, \citenamefont {Alidoust},
  \citenamefont {Liu}, \citenamefont {Wang}, \citenamefont {Lee}, \citenamefont
  {Jeng}, \citenamefont {Zhang}, \citenamefont {Yuan}, \citenamefont {Jia},
  \citenamefont {Bansil}, \citenamefont {Chou}, \citenamefont {Lin},\ and\
  \citenamefont {Hasan}}]{bian2016topological}%
  \BibitemOpen
  \bibfield  {author} {\bibinfo {author} {\bibfnamefont {G.}~\bibnamefont
  {Bian}}, \bibinfo {author} {\bibfnamefont {T.-R.}\ \bibnamefont {Chang}},
  \bibinfo {author} {\bibfnamefont {R.}~\bibnamefont {Sankar}}, \bibinfo
  {author} {\bibfnamefont {S.-Y.}\ \bibnamefont {Xu}}, \bibinfo {author}
  {\bibfnamefont {H.}~\bibnamefont {Zheng}}, \bibinfo {author} {\bibfnamefont
  {T.}~\bibnamefont {Neupert}}, \bibinfo {author} {\bibfnamefont {C.-K.}\
  \bibnamefont {Chiu}}, \bibinfo {author} {\bibfnamefont {S.-M.}\ \bibnamefont
  {Huang}}, \bibinfo {author} {\bibfnamefont {G.}~\bibnamefont {Chang}},
  \bibinfo {author} {\bibfnamefont {I.}~\bibnamefont {Belopolski}}, \bibinfo
  {author} {\bibfnamefont {D.~S.}\ \bibnamefont {Sanchez}}, \bibinfo {author}
  {\bibfnamefont {M.}~\bibnamefont {Neupane}}, \bibinfo {author} {\bibfnamefont
  {N.}~\bibnamefont {Alidoust}}, \bibinfo {author} {\bibfnamefont
  {C.}~\bibnamefont {Liu}}, \bibinfo {author} {\bibfnamefont {B.}~\bibnamefont
  {Wang}}, \bibinfo {author} {\bibfnamefont {C.-C.}\ \bibnamefont {Lee}},
  \bibinfo {author} {\bibfnamefont {H.-T.}\ \bibnamefont {Jeng}}, \bibinfo
  {author} {\bibfnamefont {C.}~\bibnamefont {Zhang}}, \bibinfo {author}
  {\bibfnamefont {Z.}~\bibnamefont {Yuan}}, \bibinfo {author} {\bibfnamefont
  {S.}~\bibnamefont {Jia}}, \bibinfo {author} {\bibfnamefont {A.}~\bibnamefont
  {Bansil}}, \bibinfo {author} {\bibfnamefont {F.}~\bibnamefont {Chou}},
  \bibinfo {author} {\bibfnamefont {H.}~\bibnamefont {Lin}}, \ and\ \bibinfo
  {author} {\bibfnamefont {M.~Z.}\ \bibnamefont {Hasan}},\ }\href {\doibase
  10.1038/ncomms10556} {\bibfield  {journal} {\bibinfo  {journal} {Nature
  Communications}\ }\textbf {\bibinfo {volume} {7}},\ \bibinfo {pages} {10556}
  (\bibinfo {year} {2016})}\BibitemShut {NoStop}%
\bibitem [{\citenamefont {Wan}\ \emph {et~al.}(2011)\citenamefont {Wan},
  \citenamefont {Turner}, \citenamefont {Vishwanath},\ and\ \citenamefont
  {Savrasov}}]{wan2011topological}%
  \BibitemOpen
  \bibfield  {author} {\bibinfo {author} {\bibfnamefont {X.}~\bibnamefont
  {Wan}}, \bibinfo {author} {\bibfnamefont {A.~M.}\ \bibnamefont {Turner}},
  \bibinfo {author} {\bibfnamefont {A.}~\bibnamefont {Vishwanath}}, \ and\
  \bibinfo {author} {\bibfnamefont {S.~Y.}\ \bibnamefont {Savrasov}},\ }\href
  {\doibase 10.1103/PhysRevB.83.205101} {\bibfield  {journal} {\bibinfo
  {journal} {Physical Review B}\ }\textbf {\bibinfo {volume} {83}},\ \bibinfo
  {pages} {205101} (\bibinfo {year} {2011})},\ \Eprint
  {http://arxiv.org/abs/1007.0016} {arXiv:1007.0016} \BibitemShut {NoStop}%
\bibitem [{\citenamefont {Xu}\ \emph {et~al.}(2015)\citenamefont {Xu},
  \citenamefont {Belopolski}, \citenamefont {Alidoust}, \citenamefont
  {Neupane}, \citenamefont {Bian}, \citenamefont {Zhang}, \citenamefont
  {Sankar}, \citenamefont {Chang}, \citenamefont {Yuan}, \citenamefont {Lee},
  \citenamefont {Huang}, \citenamefont {Zheng}, \citenamefont {Ma},
  \citenamefont {Sanchez}, \citenamefont {Wang}, \citenamefont {Bansil},
  \citenamefont {Chou}, \citenamefont {Shibayev}, \citenamefont {Lin},
  \citenamefont {Jia},\ and\ \citenamefont {Hasan}}]{xu2015discovery}%
  \BibitemOpen
  \bibfield  {author} {\bibinfo {author} {\bibfnamefont {S.-Y.}\ \bibnamefont
  {Xu}}, \bibinfo {author} {\bibfnamefont {I.}~\bibnamefont {Belopolski}},
  \bibinfo {author} {\bibfnamefont {N.}~\bibnamefont {Alidoust}}, \bibinfo
  {author} {\bibfnamefont {M.}~\bibnamefont {Neupane}}, \bibinfo {author}
  {\bibfnamefont {G.}~\bibnamefont {Bian}}, \bibinfo {author} {\bibfnamefont
  {C.}~\bibnamefont {Zhang}}, \bibinfo {author} {\bibfnamefont
  {R.}~\bibnamefont {Sankar}}, \bibinfo {author} {\bibfnamefont
  {G.}~\bibnamefont {Chang}}, \bibinfo {author} {\bibfnamefont
  {Z.}~\bibnamefont {Yuan}}, \bibinfo {author} {\bibfnamefont {C.-C.}\
  \bibnamefont {Lee}}, \bibinfo {author} {\bibfnamefont {S.-M.}\ \bibnamefont
  {Huang}}, \bibinfo {author} {\bibfnamefont {H.}~\bibnamefont {Zheng}},
  \bibinfo {author} {\bibfnamefont {J.}~\bibnamefont {Ma}}, \bibinfo {author}
  {\bibfnamefont {D.~S.}\ \bibnamefont {Sanchez}}, \bibinfo {author}
  {\bibfnamefont {B.}~\bibnamefont {Wang}}, \bibinfo {author} {\bibfnamefont
  {A.}~\bibnamefont {Bansil}}, \bibinfo {author} {\bibfnamefont
  {F.}~\bibnamefont {Chou}}, \bibinfo {author} {\bibfnamefont {P.~P.}\
  \bibnamefont {Shibayev}}, \bibinfo {author} {\bibfnamefont {H.}~\bibnamefont
  {Lin}}, \bibinfo {author} {\bibfnamefont {S.}~\bibnamefont {Jia}}, \ and\
  \bibinfo {author} {\bibfnamefont {M.~Z.}\ \bibnamefont {Hasan}},\ }\href
  {\doibase 10.1126/science.aaa9297} {\bibfield  {journal} {\bibinfo  {journal}
  {Science}\ }\textbf {\bibinfo {volume} {349}},\ \bibinfo {pages} {613}
  (\bibinfo {year} {2015})},\ \Eprint {http://arxiv.org/abs/1502.03807}
  {arXiv:1502.03807} \BibitemShut {NoStop}%
\bibitem [{\citenamefont {Lv}\ \emph {et~al.}(2015)\citenamefont {Lv},
  \citenamefont {Weng}, \citenamefont {Fu}, \citenamefont {Wang}, \citenamefont
  {Miao}, \citenamefont {Ma}, \citenamefont {Richard}, \citenamefont {Huang},
  \citenamefont {Zhao}, \citenamefont {Chen}, \citenamefont {Fang},
  \citenamefont {Dai}, \citenamefont {Qian},\ and\ \citenamefont
  {Ding}}]{lv2015experimental}%
  \BibitemOpen
  \bibfield  {author} {\bibinfo {author} {\bibfnamefont {B.~Q.}\ \bibnamefont
  {Lv}}, \bibinfo {author} {\bibfnamefont {H.~M.}\ \bibnamefont {Weng}},
  \bibinfo {author} {\bibfnamefont {B.~B.}\ \bibnamefont {Fu}}, \bibinfo
  {author} {\bibfnamefont {X.~P.}\ \bibnamefont {Wang}}, \bibinfo {author}
  {\bibfnamefont {H.}~\bibnamefont {Miao}}, \bibinfo {author} {\bibfnamefont
  {J.}~\bibnamefont {Ma}}, \bibinfo {author} {\bibfnamefont {P.}~\bibnamefont
  {Richard}}, \bibinfo {author} {\bibfnamefont {X.~C.}\ \bibnamefont {Huang}},
  \bibinfo {author} {\bibfnamefont {L.~X.}\ \bibnamefont {Zhao}}, \bibinfo
  {author} {\bibfnamefont {G.~F.}\ \bibnamefont {Chen}}, \bibinfo {author}
  {\bibfnamefont {Z.}~\bibnamefont {Fang}}, \bibinfo {author} {\bibfnamefont
  {X.}~\bibnamefont {Dai}}, \bibinfo {author} {\bibfnamefont {T.}~\bibnamefont
  {Qian}}, \ and\ \bibinfo {author} {\bibfnamefont {H.}~\bibnamefont {Ding}},\
  }\href {\doibase 10.1103/PhysRevX.5.031013} {\bibfield  {journal} {\bibinfo
  {journal} {Physical Review X}\ }\textbf {\bibinfo {volume} {5}},\ \bibinfo
  {pages} {031013} (\bibinfo {year} {2015})},\ \Eprint
  {http://arxiv.org/abs/1502.04684} {arXiv:1502.04684} \BibitemShut {NoStop}%
\bibitem [{\citenamefont {Lu}\ \emph {et~al.}(2015)\citenamefont {Lu},
  \citenamefont {Wang}, \citenamefont {Ye}, \citenamefont {Ran}, \citenamefont
  {Fu}, \citenamefont {Joannopoulos},\ and\ \citenamefont
  {Solja{\v{c}}i{\'{c}}}}]{lu2015experimental}%
  \BibitemOpen
  \bibfield  {author} {\bibinfo {author} {\bibfnamefont {L.}~\bibnamefont
  {Lu}}, \bibinfo {author} {\bibfnamefont {Z.}~\bibnamefont {Wang}}, \bibinfo
  {author} {\bibfnamefont {D.}~\bibnamefont {Ye}}, \bibinfo {author}
  {\bibfnamefont {L.}~\bibnamefont {Ran}}, \bibinfo {author} {\bibfnamefont
  {L.}~\bibnamefont {Fu}}, \bibinfo {author} {\bibfnamefont {J.~D.}\
  \bibnamefont {Joannopoulos}}, \ and\ \bibinfo {author} {\bibfnamefont
  {M.}~\bibnamefont {Solja{\v{c}}i{\'{c}}}},\ }\href {\doibase
  10.1126/science.aaa9273} {\bibfield  {journal} {\bibinfo  {journal}
  {Science}\ }\textbf {\bibinfo {volume} {349}},\ \bibinfo {pages} {622}
  (\bibinfo {year} {2015})}\BibitemShut {NoStop}%
\bibitem [{\citenamefont {Bernard}\ and\ \citenamefont
  {LeClair}(2002)}]{bernard2001}%
  \BibitemOpen
  \bibfield  {author} {\bibinfo {author} {\bibfnamefont {D.}~\bibnamefont
  {Bernard}}\ and\ \bibinfo {author} {\bibfnamefont {A.}~\bibnamefont
  {LeClair}},\ }in\ \href {\doibase 10.1007/978-94-010-0514-2_19} {\emph
  {\bibinfo {booktitle} {Statistical Field Theories}}},\ \bibinfo {editor}
  {edited by\ \bibinfo {editor} {\bibfnamefont {A.}~\bibnamefont {Cappelli}}\
  and\ \bibinfo {editor} {\bibfnamefont {G.}~\bibnamefont {Mussardo}}}\
  (\bibinfo  {publisher} {Springer},\ \bibinfo {year} {2002})\ \Eprint
  {http://arxiv.org/abs/0110649} {arXiv:0110649 [cond-mat]} \BibitemShut
  {NoStop}%
\bibitem [{\citenamefont {Magnea}(2008)}]{magnea2008random}%
  \BibitemOpen
  \bibfield  {author} {\bibinfo {author} {\bibfnamefont {U.}~\bibnamefont
  {Magnea}},\ }\href {\doibase 10.1088/1751-8113/41/4/045203} {\bibfield
  {journal} {\bibinfo  {journal} {Journal of Physics A: Mathematical and
  Theoretical}\ }\textbf {\bibinfo {volume} {41}},\ \bibinfo {pages} {045203}
  (\bibinfo {year} {2008})}\BibitemShut {NoStop}%
\bibitem [{\citenamefont {Wang}\ \emph {et~al.}(2018)\citenamefont {Wang},
  \citenamefont {Ruan},\ and\ \citenamefont {Zhang}}]{wang2018non}%
  \BibitemOpen
  \bibfield  {author} {\bibinfo {author} {\bibfnamefont {H.}~\bibnamefont
  {Wang}}, \bibinfo {author} {\bibfnamefont {J.}~\bibnamefont {Ruan}}, \ and\
  \bibinfo {author} {\bibfnamefont {H.}~\bibnamefont {Zhang}},\ }\href
  {http://arxiv.org/abs/1808.06162} {\bibfield  {journal} {\bibinfo  {journal}
  {arXiv}\ } (\bibinfo {year} {2018})},\ \Eprint
  {http://arxiv.org/abs/1808.06162} {arXiv:1808.06162} \BibitemShut {NoStop}%
\bibitem [{\citenamefont {Yan}\ \emph {et~al.}(2018)\citenamefont {Yan},
  \citenamefont {Liu}, \citenamefont {Yan}, \citenamefont {Liu}, \citenamefont
  {Chen}, \citenamefont {Wang},\ and\ \citenamefont
  {Lu}}]{yan2018experimental}%
  \BibitemOpen
  \bibfield  {author} {\bibinfo {author} {\bibfnamefont {Q.}~\bibnamefont
  {Yan}}, \bibinfo {author} {\bibfnamefont {R.}~\bibnamefont {Liu}}, \bibinfo
  {author} {\bibfnamefont {Z.}~\bibnamefont {Yan}}, \bibinfo {author}
  {\bibfnamefont {B.}~\bibnamefont {Liu}}, \bibinfo {author} {\bibfnamefont
  {H.}~\bibnamefont {Chen}}, \bibinfo {author} {\bibfnamefont {Z.}~\bibnamefont
  {Wang}}, \ and\ \bibinfo {author} {\bibfnamefont {L.}~\bibnamefont {Lu}},\
  }\href {\doibase 10.1038/s41567-017-0041-4} {\bibfield  {journal} {\bibinfo
  {journal} {Nature Physics}\ }\textbf {\bibinfo {volume} {14}},\ \bibinfo
  {pages} {461} (\bibinfo {year} {2018})}\BibitemShut {NoStop}%
\bibitem [{\citenamefont {Qi}\ \emph {et~al.}(2018)\citenamefont {Qi},
  \citenamefont {Zhang},\ and\ \citenamefont {Ge}}]{qi2018defect}%
  \BibitemOpen
  \bibfield  {author} {\bibinfo {author} {\bibfnamefont {B.}~\bibnamefont
  {Qi}}, \bibinfo {author} {\bibfnamefont {L.}~\bibnamefont {Zhang}}, \ and\
  \bibinfo {author} {\bibfnamefont {L.}~\bibnamefont {Ge}},\ }\href {\doibase
  10.1103/PhysRevLett.120.093901} {\bibfield  {journal} {\bibinfo  {journal}
  {Physical Review Letters}\ }\textbf {\bibinfo {volume} {120}},\ \bibinfo
  {pages} {093901} (\bibinfo {year} {2018})}\BibitemShut {NoStop}%
\bibitem [{\citenamefont {Altland}\ and\ \citenamefont
  {Zirnbauer}(1997)}]{altland1997nonstandard}%
  \BibitemOpen
  \bibfield  {author} {\bibinfo {author} {\bibfnamefont {A.}~\bibnamefont
  {Altland}}\ and\ \bibinfo {author} {\bibfnamefont {M.~R.}\ \bibnamefont
  {Zirnbauer}},\ }\href {\doibase 10.1103/PhysRevB.55.1142} {\bibfield
  {journal} {\bibinfo  {journal} {Physical Review B}\ }\textbf {\bibinfo
  {volume} {55}},\ \bibinfo {pages} {1142} (\bibinfo {year}
  {1997})}\BibitemShut {NoStop}%
\bibitem [{\citenamefont {Zhou}\ and\ \citenamefont
  {Lee}()}]{zhou2018periodic}%
  \BibitemOpen
  \bibfield  {author} {\bibinfo {author} {\bibfnamefont {H.}~\bibnamefont
  {Zhou}}\ and\ \bibinfo {author} {\bibfnamefont {J.~Y.}\ \bibnamefont {Lee}},\
  }\href@noop {} {\bibinfo  {journal} {(in preparation)}\ }\BibitemShut
  {NoStop}%
\bibitem [{\citenamefont {Budich}\ \emph {et~al.}(2018)\citenamefont {Budich},
  \citenamefont {Carlstr{\"{o}}m}, \citenamefont {Kunst},\ and\ \citenamefont
  {Bergholtz}}]{budich2018symmetry}%
  \BibitemOpen
\bibfield  {journal} {  }\bibfield  {author} {\bibinfo {author} {\bibfnamefont
  {J.~C.}\ \bibnamefont {Budich}}, \bibinfo {author} {\bibfnamefont
  {J.}~\bibnamefont {Carlstr{\"{o}}m}}, \bibinfo {author} {\bibfnamefont
  {F.~K.}\ \bibnamefont {Kunst}}, \ and\ \bibinfo {author} {\bibfnamefont
  {E.~J.}\ \bibnamefont {Bergholtz}},\ }\href {http://arxiv.org/abs/1810.00914}
  {\bibfield  {journal} {\bibinfo  {journal} {arXiv}\ } (\bibinfo {year}
  {2018})},\ \Eprint {http://arxiv.org/abs/1810.00914} {arXiv:1810.00914}
  \BibitemShut {NoStop}%
\bibitem [{\citenamefont {Okugawa}\ and\ \citenamefont
  {Yokoyama}(2018)}]{okugawa2018topological}%
  \BibitemOpen
  \bibfield  {author} {\bibinfo {author} {\bibfnamefont {R.}~\bibnamefont
  {Okugawa}}\ and\ \bibinfo {author} {\bibfnamefont {T.}~\bibnamefont
  {Yokoyama}},\ }\href {http://arxiv.org/abs/1810.03376} {\bibfield  {journal}
  {\bibinfo  {journal} {arXiv}\ } (\bibinfo {year} {2018})},\ \Eprint
  {http://arxiv.org/abs/1810.03376} {arXiv:1810.03376} \BibitemShut {NoStop}%
\end{thebibliography}%

\end{document}